\documentclass[longauth]{aa}
\usepackage{graphicx}
\usepackage{xcolor}

\usepackage{txfonts}
\usepackage{hyperref}

\hypersetup{%
  colorlinks=true,% hyperlinks will be black
  linkcolor={red!50!black},
  citecolor={blue!50!black},
  urlcolor={blue!80!black},  
  pdfborderstyle={/S/U/W 1}% border style will be underline of width 1pt
}

\newcommand{\teff}[0]{$T_\mathrm{eff}$}

\newcommand{\logg}[0]{$\log g$}
\newcommand{\feh}[0]{$[\mathrm{Fe/H}]$}
\newcommand{\ah}[0]{[$\mathrm{\alpha/Fe}$]}
\newcommand{\vsini}[0]{$\varv\sin{i}$}
\newcommand{\kps}[0]{$\mathrm{km}\,\mathrm{s}^{\mathrm{-1}}$}
\newcommand{\mps}[0]{$\mathrm{m}\,\mathrm{s}^{\mathrm{-1}}$}
\newcommand{\resol}[0]{$\mathcal{R}$}
\newcommand{\mas}[0]{$\mathrm{\mu as}$}
\newcommand{\grvs}[0]{$G_\mathrm{RVS}$}

\bibpunct{(}{)}{;}{a}{}{,} % to follow the A&A style

\begin{document} 

\title{A test field for Gaia}

\subtitle{Radial velocity catalogue of stars in the South Ecliptic Pole\thanks{Based on data taken with 
the VLT-UT2 of the European Southern Observatory, programmes 084.D-0427(A), 086.D-0295(A), and 
088.D-0305(A).}\fnmsep\thanks{Based on data obtained from the ESO Science Archive Facility under request number 84886.}\fnmsep\thanks{Based on data obtained with the HERMES spectrograph, installed at the Mercator Telescope, 
operated on 
the island of La Palma by the Flemish Community, at the Spanish Observatorio del Roque de los Muchachos 
of the Instituto de Astrof\'isica de Canarias and supported by the Fund for Scientific Research of Flanders (FWO),
Belgium, the Research Council of KU Leuven, Belgium, the Fonds National de la Recherche Scientifique 
(F.R.S.-FNRS), Belgium, the Royal Observatory of Belgium, the Observatoire de Gen\`eve, Switzerland and 
the Th\"uringer Landessternwarte Tautenburg, Germany.}}

\titlerunning{Radial velocity catalogue of stars at the South Ecliptic Pole}

   \author{Y.~Fr\'emat
          \inst{1}
          \and
          M.~Altmann %Heidelberg ARI
          \inst{2,3}
          \and
          E.~Pancino 
          \inst{4,5}
          \and
          C.~Soubiran 
          \inst{6}
          \and
          P.~Jofr\'e 
          \inst{6,7,8}
          \and
          Y.~Damerdji 
          \inst{9,10}
          \and
          U.~Heiter 
          \inst{11}
          \and
          F.~Royer          
          \inst{12}
          \and
		  G.~Seabroke 
          \inst{13}
          \and
		  R.~Sordo 
          \inst{14}
          \and
          S.~Blanco-Cuaresma 
          \inst{15,6}
          \and
          G.~Jasniewicz 
          \inst{16}
          \and
          C.~Martayan
          \inst{17}
          \and
          F.~Thévenin 
          \inst{18}
          \and
		  A.~Vallenari 
          \inst{14}
          \and R.~Blomme \inst{1}
          \and M.~David \inst{19}
          \and E.~Gosset \inst{10}
          \and D.~Katz \inst{12}
          \and Y.~Viala \inst{12}
          \and
          S.~Boudreault
          \inst{13,20}
          \and
          T.~Cantat-Gaudin
          \inst{14}
          \and
          A.~Lobel
          \inst{1}
          \and
          K.~Meisenheimer
          \inst{21}
          \and
          T.~Nordlander
          \inst{11}
          \and
          G.~Raskin
          \inst{22}
          \and
          P.~Royer
          \inst{22}
          \and
          J.~Zorec
          \inst{23}
          }

   \institute{Royal Observatory of Belgium, 3 avenue circulaire, B-1180 Brussels, Belgium\\ %1
              \email{yves.fremat@observatory.be} %1
         	  \and
		      Astronomisches Rechen-Institut, Zentrum f\"ur Astronomie der Universit\"at Heidelberg, M\"onchhofstr. 12-14, 69120 Heidelberg, Germany %2
			  \and		      
		      SYRTE, Observatoire de Paris, PSL Research University, CNRS, Sorbonne Universités, UPMC Univ. Paris 06, LNE, 61 avenue de l’Observatoire, 75014 Paris, France 
		      \and
              INAF - Osservatorio Astronomico di Arcetri, Largo Enrico Fermi 5, 50125 Firenze, Italy %3
		      \and
			  ASI Science Data Center, via del Politecnico snc, 00133 Rome, Italy %4
              \and
         	  LAB UMR 5804, Univ. Bordeaux - CNRS, 33270, Floirac, France %5
		      \and Institute of Astronomy, University of Cambridge, Madingley Road CB3 0HA Cambridge, UK. %6
				\and N\'ucleo de Astronom\'ia, Facultad de Ingenier\'ia, Universidad Diego Portales,  Av. Ejercito 441, Santiago, Chile %7
			  \and Centre  de  Recherche  en  astronomie,  astrophysique  et  g\'eophysique,  route  de  l’Observatoire  BP  63  Bouzareah,  16340  Algiers, Algeria %8
			  \and Space sciences, Technologies, and Astrophysics Research (STAR) Institute, Universit\'e de Li\`ege,
19c, All\'ee du 6 Ao\^ut, B-4000 Li\`ege, Belgium %9
		      \and
		      Department of Physics and Astronomy, Uppsala University, Box 516, SE-75120 Uppsala, Sweden %10
			  \and GEPI, Observatoire de Paris, CNRS, Université Paris Diderot, place Jules Janssen, 92195 Meudon Cedex, France %11
			  \and
		      Mullard Space Science Laboratory, University College London, Holmbury St. Mary, Dorking, Surrey, RH5 6NT, UK %12
		      \and
		      INAF-Osservatorio Astronomico di Padova, Vic. dell'Osservatorio 5, I-35122 Padova, Italy %13
		      \and Observatoire de Gen\`eve, Universit\'e de Gen\`eve, CH-1290 Versoix, Switzerland
              \and
         	  LUPM UMR 5299 CNRS/UM2, Universit\'e Montpellier II, CC 72, 34095, Montpellier Cedex 05, France %14 
		      \and
		      ESO - European Organisation for Astronomical Research in the Southern Hemisphere, Alonso de Cordova 3107, Vitacura, Santiago de Chile, Chile %15
                      \and
              UCA, Laboratoire Lagrange, UMR 7293, OCA, CS 34229, 06304 Nice Cedex 4, France
			  \and Universiteit Antwerpen, Onderzoeksgroep Toegepaste Wiskunde, Middelheimlaan 1, 2020 Antwerpen, Belgium %17
			\and Max Planck Institut f\"ur Sonnensystemforschung, Justus-von-
Liebig-Weg 3, 37077 G\"ottingen, Germany %18
			  \and Max-Planck Institut f\"ur Astronomie (MPIA), 69117 Heidelberg, German %19
			\and KU Leuven, Afdeling Sterrenkunde, Celestijnenlaan 200d - bus 2401, 3001 Leuven %20
			\and Sorbonne Universit\'es, UPMC Université Paris 6 et CNRS UMR 7095, Institut d’Astrophysique de Paris, F-75014 Paris, France %21
             }

\date{\today ; to appear in A\&A}

  \abstract
   {
   Gaia is a space mission currently measuring the five astrometric parameters as well as spectrophotometry 
   of at least 1 billion stars to $G = 20.7$~mag with unprecedented precision. The sixth parameter in phase 
   space --- radial velocity --- is also measured thanks to medium-resolution spectroscopy being obtained for the 
   150 million brightest stars. During the commissioning phase, two fields, one around each ecliptic pole, 
   have been repeatedly observed to assess and to improve the overall satellite performances as well as the 
   associated reduction and analysis software. A ground-based photometric and spectroscopic survey was 
   therefore initiated in 2007, and is still running in order to gather as much information as possible about
   the stars in these fields. This work is of particular interest to the validation of the Radial Velocity 
   Spectrometer (RVS) outputs.
   }
   {
   The paper presents the radial velocity measurements performed for the Southern targets in 
   the 12 - 17 $R$ magnitude range on high- to mid-resolution spectra obtained with the GIRAFFE and UVES 
   spectrographs.
   }
   {
	Comparison of the South Ecliptic Pole (SEP) GIRAFFE data to spectroscopic templates observed with 
	the HERMES (Mercator in La Palma, Spain) spectrograph allowed a first coarse characterisation of the 
	747 SEP targets. Radial velocities were then obtained by comparing the results of three different methods.   
   }
   {In this paper we present an initial overview of the targets to be found in the 1-square-degree SEP region 
   that was observed repeatedly by Gaia from its commissioning on. In our representative sample, we identified 
   1 galaxy, 6 LMC S-stars, 9 candidate chromospherically active stars, and confirmed the status of 18 LMC 
   Carbon stars. A careful study of the 3471 epoch radial velocity measurements, led us to identify 145 RV 
   constant stars with radial velocities varying by less than 1~\kps. 78 stars show significant RV 
   scatter, while nine stars show a composite spectrum. As expected the distribution of the RVs exhibits 
   2 main peaks corresponding to Galactic and LMC stars. By combining \feh\, and \logg\, estimates, 
   and RV determinations we identified 203 members of the LMC, while 51 more stars are candidate members.
   }
   {
   This is the first systematic spectroscopic characterisation of faint stars located in the SEP field. 
   During the coming years, we plan to continue our survey and gather additional high- and mid-resolution 
   data to better constrain our knowledge on key reference targets for Gaia.
   }


   \maketitle

\section{Introduction}

Gaia, ESA's astrometric satellite mission, was launched on December 19th, 2013.
It is currently obtaining the astrometry of more than 1 billion stars with 
unprecedented precision and with the ultimate goal of creating a 3-D map of 
the stellar component of our Galaxy and its surroundings. The mission will nominally 
last 5 years and is based on the principles of the earlier and successful 
Hipparcos mission.

To achieve its goals, Gaia measures the five components of the six dimensional
phase space (coordinates, parallax, and 2-D proper motion) for all objects with $G$ 
\citep[i.e., broad-band Gaia magnitude,][]{2010A&A...523A..48J} ranging from $3$ 
to $20.7$~mag \citep[Solar System bodies, $10^{\rm{9}}$ stars as well 
as several million galaxies and quasars, see][]{2012A&A...543A.100R}. 
The sixth parameter (i.e., radial velocity) is derived for the stars 
up to \grvs~$\sim 16.2$~mag \citep[i.e., Radial Velocity Spectrometer - RVS - narrow 
band magnitude, see][]{2004MNRAS.354.1223K,2011EAS....45..181C}. Final parallax errors as
small as $80$~\mas\, are expected \citep{2014A&A...566A.119L}. In addition to
accurate astrometry \citep{2011EAS....45..9}, Gaia will provide dispersed
photometry (an expected average of 70 measurements per target) in 2 wavelength 
domains (Blue Photometer, BP, and Red Photometer, RP) with a total wavelength 
coverage from $330$ to $1050$~nm and an average resolving power $\mathcal{R} \sim 20$ \citep{2010A&A...523A..48J}. 
The brighter stars will further be observed on average 40 times during 
the mission at a higher resolving power $\mathcal{R} \sim 11\,500$ by means of the RVS, 
in a spectral domain ranging from $845$ to $872$~nm known as the region of the near-IR 
\ion{Ca}{ii} triplet and of the higher members of the Paschen series. Gaia will therefore also 
classify the objects it observes, which for stars implies the determination of astrophysical 
parameters (APs). For the latest up-to-date science performance 
values, the reader may refer to the official Gaia web page 
\footnote{\url{http://www.cosmos.esa.int/web/gaia/science-performance}} and to the first 
companion paper of the Gaia Data Release 1 \citep[GDR1,][]{gdr1}.

Unlike Hipparcos, Gaia does not use an input catalogue but scans the whole sky to observe
all objects up to magnitude $G \sim 20.7$. To achieve this, it follows a scan law called 
the Nominal Scan Law (NSL) designed to 
optimise astrometric accuracy, sky coverage, and uniformity taking into account the selected orbit as 
well as other mission-related technical aspects \citep{2011EAS....45..109L}. 
In addition to the NSL, in order to perform the initial calibration of the instruments and to verify 
the high-precision astrometric performances during the commissioning phase, an Ecliptic Pole Scanning 
Law (EPSL) was introduced. This law allows, from the very beginning of the mission, to observe many 
times during a short period of time a 1-square-degree field centred on each ecliptic pole (EP), in the Southern
(SEP) and Northern (NEP) hemispheres.

%\smallskip
%\begin{tabular}{@{}rl@{}rl@{}r}
%North EP (NEP): &$\alpha=$&$18^{\rm{h}}0^{\rm{m}}0^{\rm{s}}$&$\delta=$& $+66^{\rm{o}}33^{\rm{m}}39^{\rm{s}}$\\
%South EP (SEP): &$\alpha=$&$6^{\rm{h}}0^{\rm{m}}0^{\rm{s}}$&$\delta=$& $-66^{\rm{o}}33^{\rm{m}}39^{\rm{s}}$\\
%\end{tabular}
%\smallskip

\noindent Because these are test fields, there is a continuous effort to collect as much 
knowledge as possible about them. While all literature data available for the bright stars 
were included in the Initial Gaia Source List \citep[IGSL,][]{2014A&A...570A..87S},
a photometric survey was initiated in 2007 with the MEGACAM facility on the Canadian French Hawaii 
Telescope for the North ecliptic pole and with the WFI imager on the ESO 2.2-m telescope for the South 
ecliptic pole. The aims were to collect reference data and to characterise the faintest stars of the 
EP fields between $R$ magnitudes 13 and 23. 

In a similar effort, the acquisition of reference radial velocity measurements for standard stars 
with the goal to support, and to calibrate the RVS is carried out for stars brighter than 
$V = 11$ \citep[see e.g., ][]{2010A&A...524A..10C, 2011EAS....45..195J, 2013A&A...552A..64S}. 
Other RV catalogues like 
the Geneva-Copenhagen survey \citep[GCS,][]{2004A&A...418..989N} or RAVE 
\citep[The Radial Velocity Experiment,][]{2011AJ....141..187S,2013AJ....146..134K} are 
used for large scale verification purposes. However, they also mainly include bright stars 
in the SEP, which does not permit to characterise the faint-end radial velocity performance 
of the instrument. Therefore, in the framework of the preparation of the Gaia Ecliptic Pole 
Catalogue \citep[GEPC,][]{LL:MA-015}, 747 SEP targets with $V$ magnitudes ranging from $13$ to $18$ 
were repeatedly observed (see Sect.~\ref{sect:observations}) at high- to mid-spectral resolution 
with the GIRAFFE and/or UVES spectrographs, as well as with the Wide Field Imager on the MPIA~2.2-m 
telescope (ESO) in the $B$, $V$, $R$, $I$ photometric bands. The main purpose is to characterise 
in as great detail as possible all the stars that were observed by deriving their radial velocity, 
astrophysical parameters, and chemical composition. 

The present study deals with the determination of the radial velocities on spectra obtained 
with two spectrographs, in three different wavelength ranges, and at three different
spectral resolutions. Because each RV determination method \citep{2014A&A...562A..97D} is sensitive
in a different way to instrumental, template mismatch or line blending effects, we chose to adopt 
a multi-method RV determination approach and to confront the results of the 3 independent methods.

This paper is organised as follows: the observations are described in Sect.~\ref{sect:observations},
while Sect.~\ref{sec:APhermes} explains how we perform a first characterisation of the targets.
Section \ref{sec:methods} outlines the three methods we use to derive the
radial velocities. The results are compared and analysed in Sect.~\ref{sec:results}, then
discussed in Sect.~\ref{sec:discussion}, and summarised in Sect.~\ref{sec:conclusions}.

\begin{figure}[ht]
\includegraphics[]{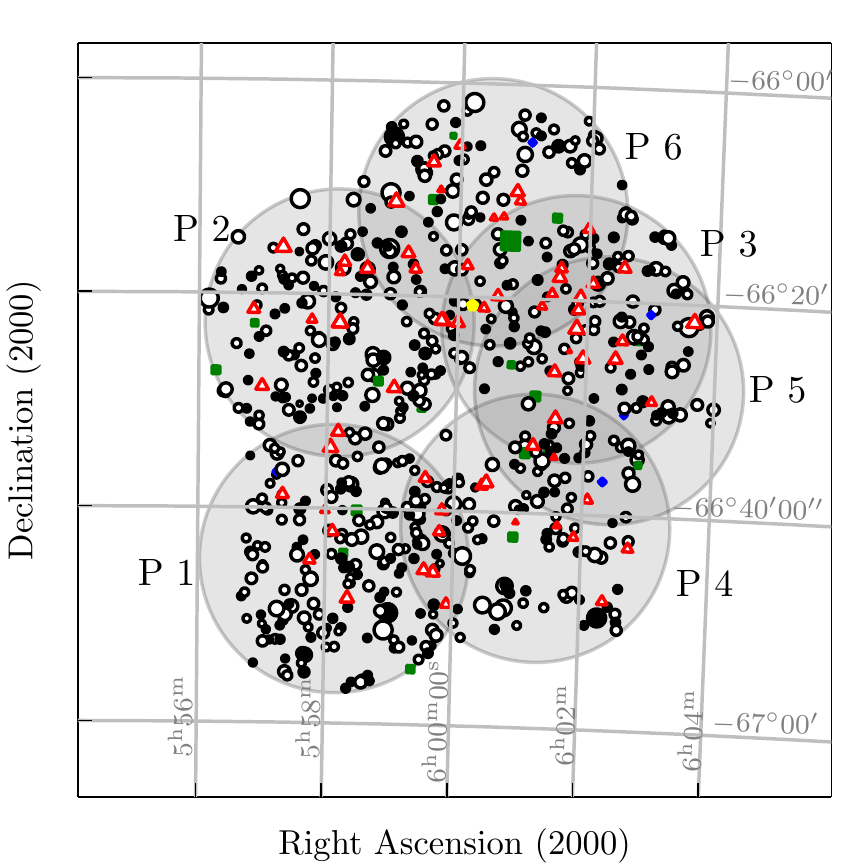}
\caption[]{South Ecliptic Pole coverage: Six different FLAMES pointings (circular shaded areas) 
were used. Symbol size is proportional to the brightness. Confirmed LMC stars (Sect.~\ref{sec:lmc}) 
are represented by filled symbols, while other targets are pictured with open symbols. UVES targets are
shown by red triangles. Carbon and S stars are marked by green squares and blue diamonds
(Sect.~\ref{sec:peculiar}), respectively, while the galaxy in our sample is located by a 
yellow pentagon. \label{fig:pointings}}
\end{figure}

\onllongtab{
\label{page:start}
\onecolumn
\tabcolsep=0.1cm
\begin{longtab}
\begin{longtable}{rrrrlrrrrr}
\caption{\label{tab:twomass} $BVRI$ photometry of South ecliptic pole stars with UVES and/or GIRAFFE data:
{The table lists the 747 observed targets with their 2MASS ID, their sequence number in the 
first version of the GEPC, and their coordinates (right ascension and declination). The coordinates
were cross-matched with the OGLE SEP catalogue of variable stars \citep{2012AcA....62..219S}. 
For the 218 matches found within $1$~arcsec, we also provide the OGLE ID as well as the angular 
distance from the tabulated coordinates. In the four last columns, we further give the $BVRI$ photometry
obtained in 2007 and 2009 at the ESO-La Silla observatory in Chile with the 2.2-m telescope and 
the WFI instrument. (The complete version of the table will be made available at the CDS.)}
}\\
\hline\hline
       & & \multicolumn{2}{c}{EPOCH 2000} & & & & & &\\
\cline{3-4}
2MASS ID & EID & RA  & DEC & OGLE ID & Ang. Dist. & $B$ & $V$  & $R$ & $I$  \\
  &   & (deg) & (deg) & & (arcsec) & (mag) & (mag) & (mag) & (mag) \\
\hline
\endfirsthead
\caption{continued.}\\
\hline\hline
       & & \multicolumn{2}{c}{EPOCH 2000} & & & & & &\\
\cline{3-4}
2MASS ID & EID & RA  & DEC & OGLE ID & Ang. Dist. & $B$ & $V$  & $R$ & $I$  \\
  &   & (deg) & (deg) & & (arcsec) & (mag) & (mag) & (mag) & (mag) \\
\hline
\endhead
\hline
\endfoot
J05560865-6621388 & 21404 & 89.036083 & -66.360722  &   &   & 16.64 & 15.96 & 15.58 & 15.40 \\
J05560908-6620343 & 21498 & 89.037875 & -66.342833  &   &   & -- & -- & -- & -- \\
J05561590-6627158 & 22969 & 89.066250 & -66.454361  & LMC563.10.14 & 0.200 & 18.30 & 16.33 & 15.06 & 13.98 \\
J05561592-6626530 & 22979 & 89.066333 & -66.448056  &   &   & 18.01 & 16.61 & 15.92 & 15.20 \\
\end{longtable}
\end{longtab}

}
 
\section{Data sample\label{sect:observations}}

\subsection{Spectroscopic observations\label{sect:data_spec}}

The spectra were obtained with ESO's VLT UT2 and the FLAMES facility \citep{2002Msngr.110....1P} 
in Medusa combined mode. In this mode, up to 132 fibres can be located on objects
feeding the GIRAFFE multi-object spectrograph, while up to 8 fibres can be placed 
and directed to the red arm of UVES.
The field of view of FLAMES is 25 arcmin in diameter. The facility has two different fibre
plates, so that while one is being observed, the other one gets configured. A small part of
the field is obstructed by the VLT guide probe\footnote{The exact area depends on 
where the VLT guide star is located in the field.}. The object positioning
was made using the FPOSS software, dedicated to set up FLAMES. A number of fibres fed into GIRAFFE
and UVES are reserved for measuring the sky background.

The GIRAFFE spectrograph offers two suites of gratings, medium resolution (''LR'', with a longer 
wavelength range) and high-resolution (''HR'', with a smaller range). We chose two gratings:
HR21, which has a resolving power of \resol$=16\,200$ and a wavelength range from $848.4$~nm to $900.1$~nm, 
and the LR2 which has a spectral domain ranging from $396.4$~nm to $456.7$~nm with \resol$=6\,400$.
For the UVES stars, we also covered the Gaia range by using the $860$~nm setup (RED860),
which uses a mosaic of 2 CCDs with a resolving power of \resol$=47\,000$. The Lower (RED--L) 
and Upper (RED--U) CCD spectra are ranging from $673$ to $853$\,nm and from $865$ to $1060$\,nm, respectively. 
UVES RED860 therefore features a gap from $853$ to $865$\,nm, right in the middle of the Gaia RVS range. 
For this reason, we reobserved most UVES stars with the HR21 grating as well. 

The HR21 setup completely encompasses the wavelength domain of the 
RVS aboard Gaia, which is the reason why this grating was chosen. It mainly contains the 
important \ion{Ca}{ii} triplet (at $849.8$~nm, $854.2$~nm, and $866.2$~nm), which is well 
suited for RV determination, as demonstrated by the RAVE survey, and which is also used 
for the estimation of metallicity \citep[e.g.,][]{2007AJ....134.1298C, 2013MNRAS.434.1681C}.
It furthermore includes the higher members of the Paschen series of hydrogen 
\citep[e.g.,][]{1995A&AS..112..475A, 1996MNRAS.279...25F} that become stronger in 
A- and B-type stars.

\begin{figure}[htbp]
\includegraphics[]{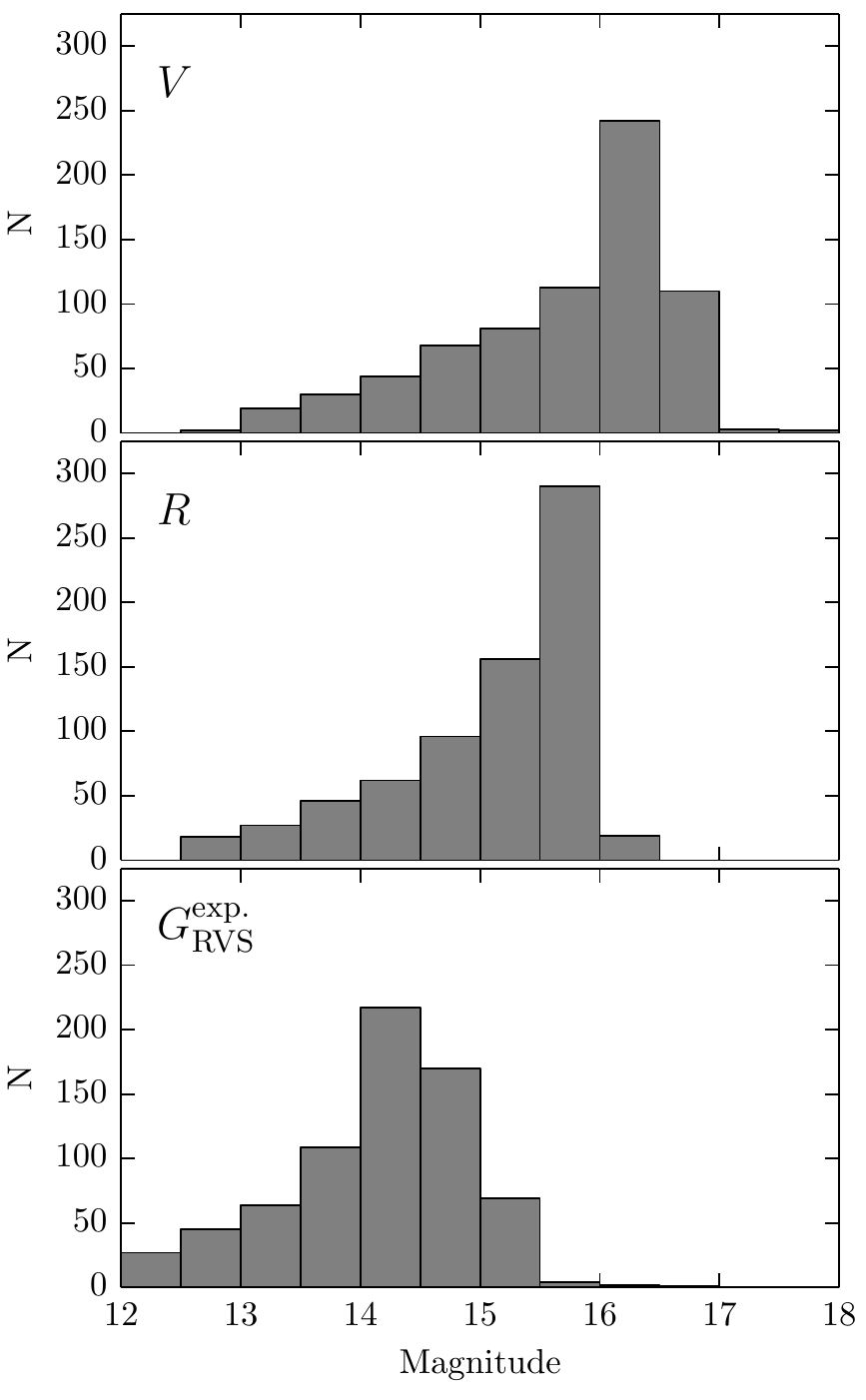}
\caption[]{Magnitude distribution of the sample: Distribution of $V$, $R$, 
and $G^\mathrm{exp.}_{\rm{RVS}}$ magnitudes of the stars observed with the GIRAFFE and 
UVES spectrographs. $G^\mathrm{exp.}_\mathrm{RVS}$ is the expected RVS narrow band 
magnitude that was estimated from the relations provided by \citet{2010A&A...523A..48J}.
\label{fig:magnitudes}}
\end{figure}

To ease the exposure time computation and to carry-out the observations in the most
efficient way, the GIRAFFE stars were selected randomly within two magnitude bins (the magnitude 
distributions of the sample are provided in Fig.~\ref{fig:magnitudes}), one brighter
than $R\sim15$~mag, and one ranging from $R\sim 15~\mathrm{to}~16.5$~mag, which approximately 
gives the lower brightness limit for the RVS instrument of Gaia. The UVES stars were chosen 
to be brighter than the GIRAFFE stars belonging to the same exposure, since the higher resolution of 
this instrument demands more light to reach 
a sufficiently high SNR level. The FPOSS positioning software allows for several modes
of fibre allocation. Because in this magnitude range our field was not crowded, the allocation
mode was not critical. Usually we could allocate fibres to most target stars, with some fibres 
to spare. This especially holds true for those exposures in which the brighter stars were covered. 
Nonetheless fibre conflicts did occur, and thus not all of the selected stars could be observed. 

We used six different pointings (see Fig.~\ref{fig:pointings}). As some of these circular 
fields do overlap, there are a number of stars present in more than one pointing. 
The stars within a field and within a magnitude bin were selected 
randomly, in such a way as to ensure a maximum of targets being allocated with fibres.
This further also ensures 
an unbiased stellar sample  (except for the magnitude limits), which is representative of the 
SEP-field in the aforementioned magnitude ranges. In total, 747 objects have been observed 
and listed in Table~\ref{tab:twomass} with their 2MASS ID and their sequence number in the 
first version of the GEPC (EID). For the sake of convenience, we will hereafter only designate 
the stars by their EID. Table~\ref{tab:twomass} further provides the right ascension and
the declination, which were cross-matched with the OGLE SEP catalogue of variable stars 
\citep{2012AcA....62..219S}. We found 218 matches within $1$~arcsec, and provide their 
OGLE ID as well as the angular distance from the tabulated coordinates in the 2 next 
columns of the same table. 

The observations were done over 4 semesters (ESO P82, 84, 86, 
88\footnote{The P82 campaign did not deliver any data.})
in service mode. ESO-Observing blocks are generally restricted to last not more 
than 1 hour in total, therefore the observations were split into several exposures. 
Table~\ref{tab:journal} is an overview of the observations. It gives the observing
block ID (OB), the modified Julian date (MJD), GIRAFFE setup, plate number, exposure time, 
{period of observation,} pointing direction (see also Fig.\,\ref{fig:pointings}), and median per 
sample SNR value, 
respectively.

\onllongtab{
\begin{table*}[htbp]
\caption{\label{tab:journal} GIRAFFE observations log (The complete version of the table will be made available at the CDS.)}
\center
\tabcolsep=0.1cm
\begin{tabular}{rrlrrrlr}
\hline\hline
OB ID & MJD & Setup & Plate & texp  & Period & Pointing & SNR \\
      &     &       &       &  (s)  &  & & \\
\hline
  389193  &  55185.34904  &  LR2   &   1  &  2000  & P84 &  P1  &  9 \\
  389179  &  55202.23191  &  HR21  &   2  &  2700  & P84 &  P3  &  7 \\
  389178  &  55203.31727  &  HR21  &   2  &  2200  & P84 &  P3  &  13 \\
 \hline
\end{tabular}
\tablefoot{
{The table provides the observing block ID (OB), the modified Julian date (MJD), 
GIRAFFE setup, plate number, exposure time, period of observation, 
pointing direction (see also Fig.\,\ref{fig:pointings}), and median per sample SNR value.}
}
\end{table*}

}

\subsubsection{Data reduction\label{sect:reduction_spec}}

The raw frames of the GIRAFFE observations have been reduced by us using the 
ESO GIRAFFE pipeline (release 2.12.1). Before running the localisation recipe, automatic 
identification of missing fibres has been performed on the raw flat-field frames and 
fed as input to the recipe. One-dimensional spectra were extracted, for both flat-field 
and science frames, using the optimal extraction method. 

{
Sky lines are spread over the HR21 wavelength domain, and need
to be removed. To do that we adopted a methodology inspired by the one described 
in \citet{2008MNRAS.383..183B}. We applied k-sigma clipping to scale the sky lines of
the median sky spectrum to those extracted from the object spectrum. The rescaled
median sky spectrum was then subtracted from the object spectrum and the result smoothed. 
When no sky fibres were available, we used the median sky of the closest exposure. For the 
final continuum normalisation or correction, we proceeded as explained in 
Sect.~\ref{sec:correlation by templates}.

It is worth adding that we do not shift the object spectrum to place it in the framework 
of the sky lines as is done by \citet{2008MNRAS.383..183B}. The clipped out sky lines were 
rather used to evaluate the median shift and its dispersion by comparison to the median 
sky lines of a given night ($\mathrm{MJD} = 55\,271.13419$, see Table~\ref{tab:journal}).
In this way, the median sky line shift 
obtained for the dataset is $+0.19$~\kps\,with a dispersion \citep[measured as half the interquantile range 
from 15.87\% to 84.13\%, see eq.19 in][]{2014A&A...562A..97D}
of $0.25$~\kps\,(P84 over 973 spectra: $0.13\pm0.24$~\kps; P86 over 1065 spectra: $0.25\pm0.30$~\kps; P88
over 47 spectra: $-0.16\pm0.24$~\kps). This definition for dispersion will be used throughout the text.
The scatter reflects the mean precision and stability level of the wavelength calibration 
over the 3 years of observation.  

Additionally there are a few telluric-lines present at the very
red edge of the same wavelength range (mainly from $890$ to $900$~nm) that we cross-correlated 
to a synthetic spectrum of the Earth atmosphere transmission generated by the TAPAS web 
server \citep{2014A&A...564A..46B}. The median shift of the telluric-lines over 
the 3 periods we observed is $+0.49\pm0.74$~\kps, which is consistent with the 
value obtained for each one individually (P84: $+0.52\pm0.67$~\kps; 
P86: $+0.47\pm0.81$~\kps; P88: $+0.26\pm0.40$~\kps).
}

UVES spectra were reduced with the UVES pipeline \citep{2000Msngr.101...31B,2004Msngr.118....8M} 
which performs bias and flat-field correction (this last step also corrects for the 
different fibre transmission), spectra tracing, optimal extraction, and wavelength calibration. 
We generally used two of the eight available UVES fibres to obtain sky spectra. After
comparing the sky spectra, to verify that they were compatible with each other and
showed no artifacts or clear contamination from faint stars, we created a master
sky spectrum as an average of the two sky spectra, and subtracted it from all
scientific spectra of the same pointing. The sky-subtracted UVES spectra were
normalised by their continuum shape with low order polynomials.
The telluric-line shifts relative to the TAPAS data were then measured by using the
same approach as for GIRAFFE data over all 3 observed periods, which provided 
a median shift of $-0.64\pm0.45$~\kps\,(P84 over 60 spectra: $-0.68\pm0.47$ \kps; 
P86 over 119 spectra: $-0.57\pm0.43$ \kps; P88
over 34 spectra: $-0.76\pm0.42$ \kps). {When considering RED--L and RED--U measurements
independently, the median of the differences, $\overline{\Delta \mathrm{RV}_\mathrm{UL}} = \mathrm{RV}_\mathrm{U} - \mathrm{RV}_\mathrm{L}$, is $0.08 \pm 0.19$~\kps.} 

Because the telluric-lines are well distributed over the observed wavelength domain, and that
we also use ESO archive data (see Sect.~\ref{sec:comparisonwcatalogue}) 
spread over a long period of time, all RV values measured on the 
UVES spectra were brought into the framework of the Earth atmosphere absorption lines.

\begin{figure}[htbp]
\includegraphics[]{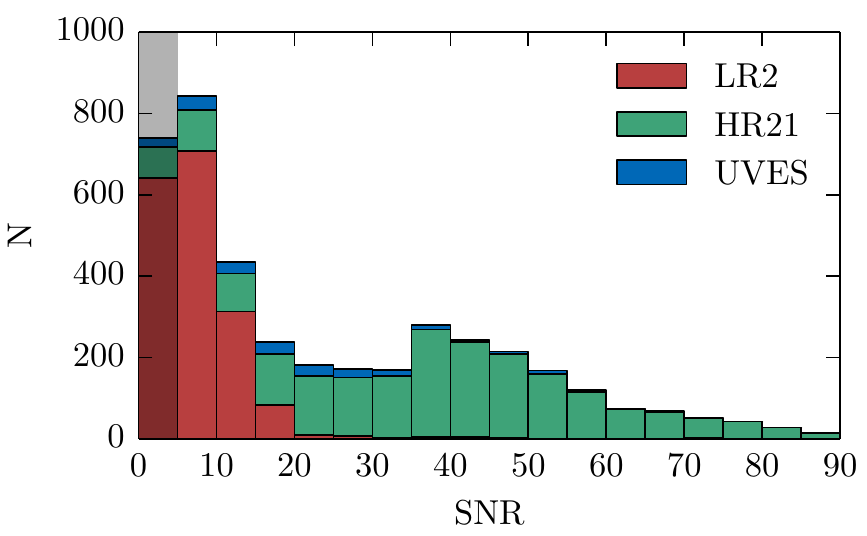}
\caption[]{{Stacked histograms of the SNR of all observations}: Distribution (bin width = 5) 
of the SNR per sample in the GIRAFFE and UVES spectra. 
Spectra with SNR\,$<$\,5 (gray shaded region)
have only been considered for stellar classification {to obtain a co-added spectrum}.\label{fig:SNR}}
\end{figure}

As can be seen in the stacked histogram of Fig.~\ref{fig:SNR},
due to the nature of the SEP stellar population mainly made of cooler stars, 
HR21 observations generally achieved higher SNR.
In total, 4129 spectra were reduced from which 3471 (190 UVES, 1167 LR2, 2114 HR21) 
had an $\mathrm{SNR} \ge 5$.   
Therefore, while 747 targets have been observed,
only 724 have epoch spectra of sufficiently high quality to derive epoch radial velocities
and discuss their variability. {Despite of this, we included the lower quality 
data to produce the co-added spectra we use to determine the best template as
explained in Sect.~\ref{sec:templateoptimization}.}

\subsection{Photometric observations\label{sect:data_phot}}

The photometric data were obtained in 2007 and 2009 at the ESO-La Silla observatory
in Chile with the 2.2-m telescope and the WFI instrument. The November 2007 data was 
incomplete and did not allow for photometric calibration. Both issues were remedied with the 
second run in January 2009. The data was reduced using MIDAS and the MPIAphot \citep{mpiaphot86}
suite. Calibration was done using Landolt standards \citep{1992AJ....104..340L}. While the photometry 
and astrometry of the SEP and NEP will be published in a separate paper, we provide the $BVRI$ 
magnitudes in the 4 last columns of Table~\ref{tab:twomass} as these values were used 
to initiate the template optimisation procedure (Sect.~\ref{sec:templateoptimization})
as well as to estimate the expected brightness coverage in terms of \grvs\, magnitude 
\citep[see conversion relations in][]{2010A&A...523A..48J}.
$R$, $I$, and expected \grvs~magnitude distributions are plotted in Fig.\,\ref{fig:magnitudes}.

\section{First insight into the data using HERMES spectra\label{sec:APhermes}}

%In a situation of randomly selected targets, 

Because the targets were selected at random and that there is
not much information available on such faint stars, we decided
to perform a preliminary target characterisation based on the use 
of HERMES reference observations.

\begin{figure}[htbp]
\includegraphics[]{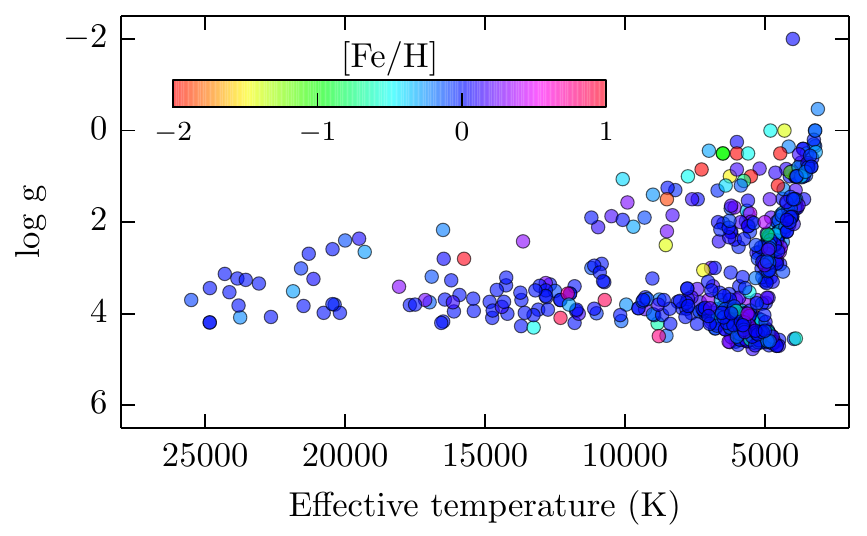}
\caption[]{Parameter space coverage of the HERMES reference spectra: \logg\, is reported as a function of \teff,
while a colour code is used to picture the published \feh. When \feh\, is unknown, the value
is assumed to be zero.
\label{fig:hermesaprange}}
\end{figure}

Since the beginning of the exploitation of the high-resolution HERMES spectrograph 
\citep{2011A&A...526A..69R} in 2009 on the Mercator telescope (Roque de los Muchachos
Observatory on La Palma, Canary Islands, Spain), a large volume of data has been
acquired. One of the observing projects is running without interruption
since the beginning of operations (late 2009). It is
conceived as a fill-in program (PI: P.~Royer) and consists in the
acquisition of high SNR data across the HR diagram. We performed
a first empirical classification of the GIRAFFE data by
systematically comparing the SEP spectra to a subsample of the
HERMES library. Our sample of HERMES template stars was
built by selecting 641 stars having a spectrum with $\mathrm{SNR} > 100$,
with known astrophysical parameters (\teff\, and \logg), and 
mainly single lined. The \logg\, vs \teff\, coverage of the
library is plotted in Fig.~\ref{fig:hermesaprange}, with a colour code 
that matches the published \feh. Each spectrum was convolved with a Gaussian
LSF to match the spectroscopic resolution of the LR2 and
HR21 settings, then compared to the available GIRAFFE spectra
by adopting a cross-correlation technique. These comparisons
have been performed in the 2 settings independently, in order to
potentially detect binary components with different colours. The
identification and parameters corresponding to the most similar
HERMES spectrum (i.e., with the highest correlation peak)
are provided in Table~\ref{tab:hermestemplates} in the following order: 
star’s EID, explored GIRAFFE Setup (GS), and for each GS the ID of the HERMES
reference target with the closest spectrum, its \teff, \logg,
\feh, and \vsini. The origin of the spectral type and parameters
are given between brackets. In cases where no literature
reference is provided for the spectral type, the classification is
the one found in the basic information tab in the Simbad database (CDS),
while the same situation for the astrophysical parameters means that 
the values were estimated
using the available Str\"omgren photometry as well as the 
\citet{1985MNRAS.217..305M} calibration updated by \citet{napiwotzki}.
We observe that LR2 and HR21 \teff\, determinations are generally in good
agreement, as the median difference and dispersion are of the order of $244 \pm 236$~K.
When the \teff\, in LR2 and HR21 wavelength domains are different
by more than $1000$~K, the target is highlighted in bold font.

\onllongtab{

\onecolumn
\tabcolsep=0.1cm
\begin{longtab}
\begin{longtable}{rlllcrrrcrc}
\caption{\label{tab:hermestemplates} SEP targets and characteristics of the corresponding closest HERMES spectrum:
{
The table provides the star’s EID, explored GIRAFFE Setup (GS), and for each GS the ID 
of the HERMES reference target with the closest spectrum, its \teff, \logg,
\feh, and \vsini. The origin of the spectral type and parameters are given between brackets and in the
table footer. When no literature reference is provided for the spectral type, the classification is
the one found in the basic information tab in the Simbad database (CDS),
while the same situation for the astrophysical parameters means that 
the values were estimated using the available Str\"omgren photometry as well as the 
\citet{1985MNRAS.217..305M} calibration updated by \citet{napiwotzki}.
When the \teff\, in LR2 and HR21 wavelength domains are different
by more than $1000$~K, the target is highlighted in bold font. (The complete version of the table will be made available at the CDS.)
}
}\\
\hline\hline
       & & \multicolumn{9}{c}{Closest HERMES spectrum}\\
\cline{3-11}
EID & GS & ID      & Sp. Type & & \teff & \logg & \feh & & \vsini & \\
\hline
\endfirsthead
\caption{continued.}\\
\hline\hline
       & & \multicolumn{9}{c}{Closest HERMES spectrum}\\
\cline{3-11}
EID & GS & ID      & Sp. Type & & \teff & \logg & \feh & & \vsini & \\
\hline
\endhead
\hline
\endfoot
21404 & LR2 & HD 18757  & G1.5 V & (1)  &   5741  &   4.30  &   -0.25  & (2)  &   0  & (2)  \\ & HR21 & HD 18757  & G1.5 V & (1)  &   5741  &   4.30  &   -0.25  & (2)  &   0  & (2)  \\
21498 & LR2 & HD 115604  & F3III & (3)  &   6855  &   3.40  &    0.18  & (4)  &   0  & (4)  \\ & HR21 & HD 23230  & F5IIvar & (25)  &   6675  &   2.00  &    0.04  & (55)  &  44  & (48)  \\
22969 & LR2 & HIP 86162  & M3 & (5)  &   3583  & 4.86 &   0.00  & (30)  &  &  \\ & HR21 & HD 92839  & C 5 II & (36)  &   3623  & 4.80 &    -0.10  & (24)  &  &  \\
22979 & LR2 & HD 20618  & G8IV & (6)  &   5093  &   3.20  &  & (4)  &   2  & (4)  \\ & HR21 & HD 159222  & G1 V & (1)  &   5788  &   4.25  &  & (1)  &   3  & (14)  \\
23240 & LR2 & HD 42995  & M3III & (7)  &   3600  &   1.50  &    0.04  & (8)  &  &  \\
\end{longtable}
\tablebib{
(1) \citet{2003AJ....126.2048G};
(2) \citet{2012A&A...547A.106M};
(3) \citet{2009MNRAS.396.1895C};
(4) \citet{2008AJ....135..209M};
(5) \citet{2012ApJ...748...93R};
(6) \citet{1969AJ.....74..916H};
(7) \citet{1962RGOB...51...79E};
(8) \citet{1998A&A...338..623M};
(9) \citet{1967AJ.....72.1334C};
(10) \citet{2011AJ....141...90L};
(11) \citet{2011A&A...531A.165P};
(12) \citet{2012A&A...547A..13T};
(13) \citet{2004A&A...420..183A};
(14) \citet{2010A&A...520A..79M};
(15) \citet{2009ApJ...697..544S};
(16) \citet{2003MNRAS.340..304R};
(17) \citet{2010MNRAS.407.1657H};
(18) \citet{1961POHP....5...54F};
(19) \citet{1979ApJ...234..538A};
(20) \citet{1986ApJ...311..843S};
(21) \citet{1990ApJ...354..310B};
(22) \citet{2009A&A...493.1099S};
(23) \citet{2012MNRAS.423..122B};
(24) \citet{2009A&A...508.1313F};
(25) \citet{1960MNRAS.120..287G};
(26) \citet{1995ApJS...99..659V};
(27) \citet{2012AJ....144...20A};
(28) \citet{2007A&A...463..671R};
(29) \citet{1973ARA&A..11...29M};
(30) \citet{2008A&A...480...91S};
(31) \citet{2011ApJ...736...87K};
(32) \citet{1972ApJ...175..453M};
(33) \citet{2013A&A...551A..30B};
(34) \citet{2007A&A...463.1071K};
(35) \citet{2011AJ....142..136L};
(36) \citet{1971ApJ...167..521R};
(37) \citet{2011A&A...530A.138C};
(38) \citet{2001MNRAS.328...45M};
(39) \citet{2007PASJ...59..335T};
(40) \citet{2007AJ....133.2524W};
(41) \citet{1968ApJS...17..371L};
(42) \citet{2010ApJ...722..605H};
(43) \citet{2002ApJ...573..359A};
(44) \citet{2001ApJ...553..754B};
(45) \citet{2005PASJ...57...13T};
(46) \citet{2012A&A...543A.160T};
(47) \citet{1998ApJS..119...83P};
(48) \citet{1970CoKwa.189....0U};
(49) \citet{2012A&A...538A.143K};
(50) \citet{1951ApJ...113..304B};
(51) \citet{2011ApJ...742...54H};
(52) \citet{1958MNRAS.118..154E};
(53) \citet{2006ApJ...646..505B};
(54) \citet{1991ApJS...77..515L};
(55) \citet{1995AJ....110.2425L};
(56) \citet{2010ApJ...720.1290G};
(57) \citet{2012ApJ...761..161S};
(58) \citet{1956PDDO....2..105H};
(59) \citet{1990ApJS...72..387S};
(61) \citet{1990AJ.....99.1961F};
(62) \citet{2006AJ....132..161G};
(63) \citet{1995A&A...294..536H};
(64) \citet{1990ApJS...74.1075M};
(65) \citet{1976AJ.....81..245E};
(66) \citet{1969AJ.....74..375C};
(67) \citet{2011A&A...533A..99G};
(68) \citet{1960ApJ...131..330K};
(69) \citet{2007ApJ...668.1165M};
(70) \citet{2009A&A...498..961R};
(71) \citet{1996A&A...314..191G};
(72) \citet{1970ApJS...19..281B};
(73) \citet{2000A&A...363..692T};
(74) \citet{1957PASP...69..326B};
(75) \citet{2000ApJ...542..978B};
(76) \citet{2002A&A...393..897R};
(77) \citet{1954ApJ...120..484K};
(78) \citet{1974ApJ...193..113C};
(79) \citet{2001AJ....121.2148G};
(80) \citet{1988IAUS..132..449C};
(81) \citet{1998A&AS..129..237F}.
}
\end{longtab}

}

\section{RV determination methods\label{sec:methods}}

We chose to adopt a multi-method approach based on the confrontation and
combination of three independent RV determination
procedures described in the following subsections.

\subsection{Pearson correlation\label{sec:correlation by templates}}

The correlation of observed spectra with theoretical ones is performed by computing the Pearson
correlation coefficient \citep[][and references therein]{2014A&A...562A..97D} for different
radial velocity shifts. To derive the star's RV, we have reduced the step around the top of 
the Pearson Correlation Function (PCF) to 1~\kps, then we combined the solution of 2 parabolic fits 
through 3 and 4 points taken on both sides of the maximum as described in \citet{1995A&AS..111..183D}.
All the data manipulation, such as Doppler shifting and
wavelength resampling, is applied on the template. Intrinsic errors on RV are deduced 
from the maximum-likelihood theory \citep{2003MNRAS.342.1291Z}. We used a {\tt Fortran} procedure which we
named {\tt PCOR} (Pearson Correlation) to perform these operations.
%%%%%%%%%%%%%%%%%%%%
{
To reduce the impact of possible mismatches at the spectra edges (e.g., due to normalisation) 
as well as to limit the blends with telluric lines, the correlation was computed in wavelength domains
that exclude such features. In practice, we therefore only considered the spectral ranges listed 
in Table~\ref{tab:pcor.regions} with their corresponding instrument and observing mode/region, as well as their 
beginning ($\lambda_\mathrm{beg.}$), and end ($\lambda_\mathrm{end}$) wavelengths. 
In the case of the UVES data and for further comparisons with the 2 other methods, the radial velocity we kept 
is the median over the 5 available domains after 
correction of the telluric 
line shift and after having filtered out potential outliers thanks to Chauvenet's criterion. 
For information purposes, we note that if a separate value for the UVES RED--L and RED--U 
spectra is computed, we obtain a median difference, 
$\overline{\Delta \mathrm{RV}_\mathrm{UL}} = \mathrm{RV}_\mathrm{U} - \mathrm{RV}_\mathrm{L}$, 
between both estimates of 
$0.14 \pm 0.26$~\kps\,(where the error is the dispersion obtained for 158 spectra).

\begin{table}
\caption{\label{tab:pcor.regions} Wavelength ranges taken into account by the {\tt PCOR} (Pearson Correlation) 
procedure while deriving the stellar RVs}

\begin{tabular}{lll}
\hline\hline
 Spectrograph--region & $\lambda_\mathrm{beg.}$ & $\lambda_\mathrm{end}$ \\
                   & (nm) & (nm) \\
\hline
GIRAFFE--LR2 & 400. & 450. \\
GIRAFFE--HR21 & 849. & 875. \\
UVES--REDL & 676.1 & 686.0 \\
          & 694.0 & 708.0 \\
          & 737.4 & 756.4 \\
          & 773.5 & 811.3 \\
UVES--REDU & 868.0 & 889.0 \\ 
\hline
\end{tabular}
\end{table}

}

%%%%%%%%%%%%%%%%%%%%%%%%

Besides random and calibration errors, one possible source of systematic error on the measurement 
of RVs is the mismatch between the observations and the synthetic spectra used to construct the 
templates. The main mismatch problems originate from the use of inappropriate stellar atmosphere 
parameters, the application of inappropriate line broadening to account for the instrumental 
LSF or for stellar rotation, as well as imperfect or inconsistent (compared to the template) 
continuum normalisation of the observations. In order to reduce the impact of these errors, we adopted 
different spectral libraries that cover as closely as possible the expected range 
of atmospheric parameters, 
and applied a multi-step RV determination aimed to iteratively improve the choice and construction 
of the template spectrum. 

\subsubsection{Synthetic spectra libraries}

Depending on the effective temperature range, different publicly available 
libraries of spectra were considered. To represent stars cooler than 3500~K, we adopted
%\footnote{http://phoenix.ens-lyon.fr/Grids/BT-Settl/CIFIST2011/SPECTRA/}
the BT-Settl {\sc PHOENIX} grid from \citet[][]{2013A&A...556A..15R}. From 3500~K to 5500~K, 
the models computed by
%\footnote{http://www.mpa-garching.mpg.de/PUBLICATIONS/DATA/SYNTHSTELLIB/synthetic\_stellar\_spectra.html}
\citet{2005A&A...443..735C} were taken, while 
from 5500~K to 15000~K, {\sc MARCS} \citep{2008A&A...486..951G} and {\sc ATLAS} \citep{2004astro.ph..5087C} model
spectra were downloaded from the {\sc Pollux} database \citep{2010A&A...516A..13P}. For higher effective 
temperatures, we used 
%\footnote{http://nova.astro.umd.edu/Tlusty2002/tlusty-frames-BS06.html}
the BSTAR flux grids computed 
by \citet{2007ApJS..169...83L}. All these 4593 spectra were convolved with a Gaussian instrument
profile at the appropriate spectral resolution.

\begin{figure}[htbp]
\begin{center}
\includegraphics[width=\columnwidth]{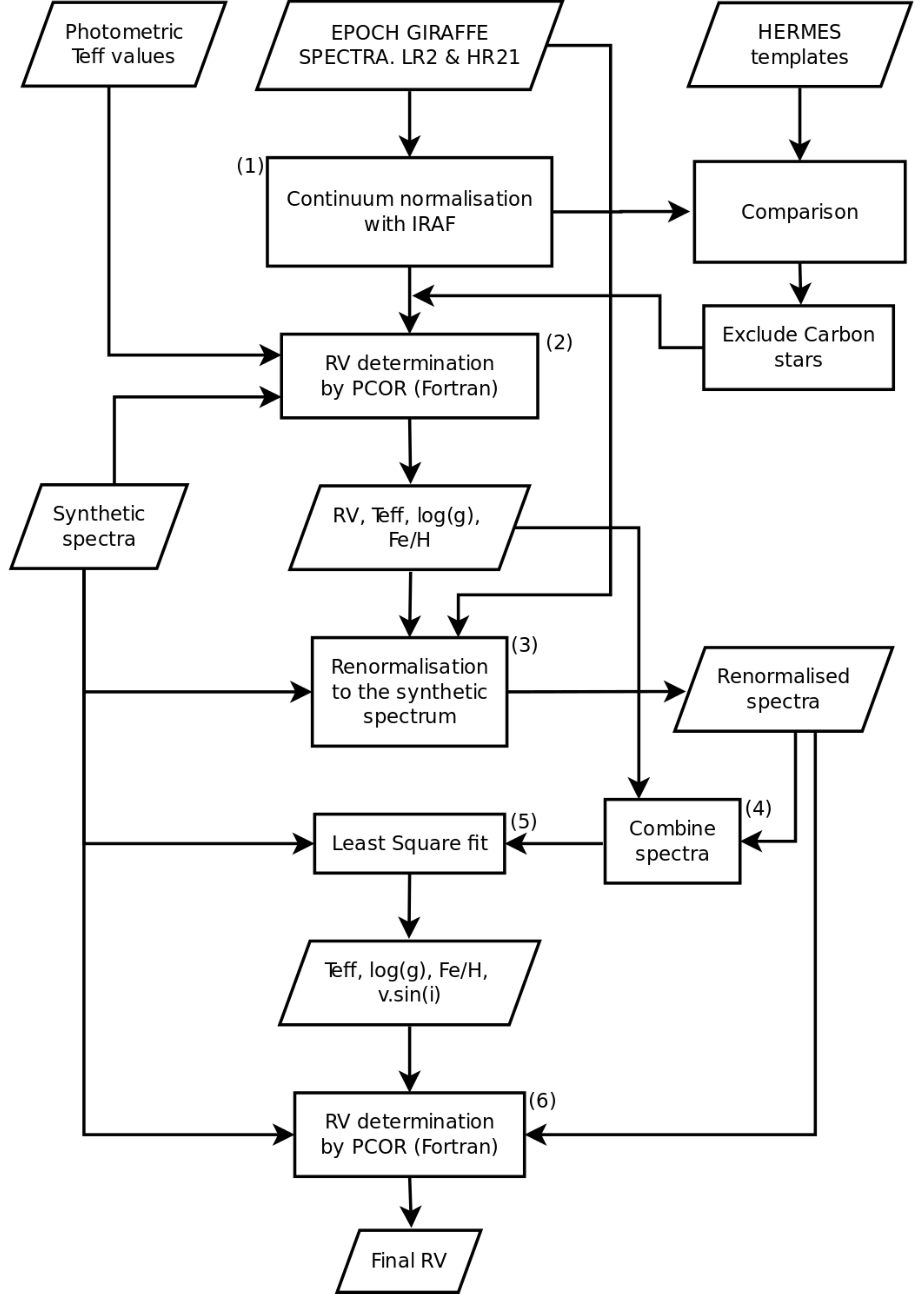}
\caption[]{Template optimisation with Pearson Correlation: outline of the procedure described in Sect.~\ref{sec:templateoptimization}\label{fig:spccf}}
\end{center}
\end{figure}

\subsubsection{Template optimisation and RV determination\label{sec:templateoptimization}}

The outline of the procedure which combines various {\tt python} scripts and one
{\tt Fortran} program is drawn in Fig.\ref{fig:spccf}. 
In a first step {(1)}, the spectra 
are automatically normalised by means of {\tt IRAF}'s {\tt continuum} task. 
At this point, the carbon stars that we have identified by comparison with 
the HERMES data (see Sect.~\ref{sec:APhermes}) are excluded, as
no corresponding template spectra are available in our libraries.
A first estimate of the radial velocities is done in the next step {(2)}, 
while the best template is chosen by maximisation of 
the PCF among a subset of spectra 
chosen within 500 K of a first estimate of the 
effective temperature. We obtained this estimate by applying where possible the calibrations of
\citet{2011ApJS..193....1W} and, to a lesser extent, \citet{2009A&A...497..497G} 
to the $BVRI$ photometry (Sect.~\ref{sect:data_phot})
supplemented by $JHK$ magnitudes available in the 2MASS catalogue.
When no initial value is available, the choice is made among
a predefined list of templates spread over the complete AP space.
Accounting for the different metallicity and 
surface gravity values, it represents 50 to 200 templates to be applied on all 
HR21, LR2, and UVES spectra 
available for a given target. 
Having a first estimate of RV as well as of the astrophysical parameters,
all the spectra are renormalised {(3)} using the currently available optimised template. 
This part of the work is
performed by fitting a polynomial of second (HR21) and third (LR2) order through the ratio
of the observed and synthetic spectra, then by applying this normalisation function to 
the observations. For the GIRAFFE data, the choice of the polynomial was done by performing the operation
manually on the spectra of a few targets with different spectral types and observed through the 
3 periods (P84, P86, and P88), while UVES data were considered individually. 
{
All the normalised observed spectra (including those with $\mathrm{SNR} < 5$) of a given star in a given
setup were then corrected for the RV shift derived from step {(2)} and 
co-added (we computed an average weighted by their SNR) {(4)}.
}
In step {(5)}, all 
available synthetic spectra are compared to the combined spectrum in order to select the 
final best template (in terms of sum of square differences) as well as the additional potential 
line broadening term (we will refer to it as \vsini, but it may also include other broadening mechanisms 
difficult to disentangle from each other such as macro-turbulence). The last RV measurement is performed 
by applying the final best 
template to the newly normalised observed spectra {(6)}. The stars' EID, their astrophysical 
parameters (\teff, \logg, \feh, and \ah) and, where needed, the \vsini\, of the final 
templates are provided in Table~\ref{tab:parameters}. 

\onllongtab{
\onecolumn
\tabcolsep=0.1cm
\begin{longtab}
\begin{longtable}{rrrrrr}
\caption{\label{tab:parameters} Stellar parameters {(\teff, \logg, \feh, \ah, and -- where needed -- the \vsini)} of the final best 
template used with the {\tt PCOR} procedure (The complete version of the table will be made available at the CDS.)}\\
\hline\hline
 EID &  \teff & \logg & \feh & \ah &  \vsini \\
   &  (K) &  & (dex) & (dex) &  (\kps) \\
\hline
\endfirsthead
\caption{continued.}\\
\hline\hline
 EID &  \teff & \logg & \feh & \ah &  \vsini \\
   &  (K) &  & (dex) & (dex) &  (\kps) \\
\hline
\endhead
\hline
\endfoot
 21404 &  5500 &   4.0 & $-$0.25 &   0.1 &    \\
 21498 &   6500  &   4.0 &  0.25 &   0.0 &   20   \\
 22969 & \multicolumn{5}{c}{C star}\\
 22979 &  4750 &   2.0 &  0.00 &   0.4 &    \\
\end{longtable}
\end{longtab}

}

\subsection{Linelist cross-correlation with {\tt DAOSPEC}\label{sec:daospec}}

{\tt DAOSPEC} \citep{daospec} is an automated {\tt Fortran} program to measure
equivalent widths of absorption lines in high-quality stellar spectra (roughly,
spectra with resolving power \resol$\gtrsim$10\,000 and SNR$\gtrsim$30). It
cross-matches line centres found in the spectrum (first by a
second-derivative  filter, and later refined by Gaussian fits) with a list of
laboratory wavelengths provided by the user. 
{\tt DAOSPEC} is generally used to measure EWs, but an RV determination based 
on the cross-correlation is also provided. The quality of the resulting RV is 
expected to be comparable to other methods, albeit with a slightly larger scatter. 
This is caused by the fact that only the line centre information is used, compared 
to cross-correlation methods that use all pixels in a spectrum or in a set of pre-defined regions.
In the context of this
paper, we also tested {\tt DAOSPEC} outside its validity regime, i.e., 
LR2 spectra at
low SNR (\resol\,$<$20\,000 and SNR generally below 30). As a result, we saw that
not only the determined RVs tend to be more scattered than those obtained with
template or mask cross-correlation spectra, but there was a systematic offset,
that increases slightly as SNR decreases, \teff\, decreases, or \feh\,
increases. This confirms that {\tt DAOSPEC} can only be used within the validity
limits stated in \citet{daospec} in terms of SNR and spectral resolution. 
Within its validity limits, however, {\tt DAOSPEC} produces results that
are well comparable to those obtained with cross-correlation
methods, and can thus be used to produce scientifically useful RV measurements
as a direct by-product of any EW-based abundance analysis. 

To test {\tt DAOSPEC} on the spectra presented in Section~\ref{sect:data_spec}, we
prepared a raw linelist with the help of a dozen synthetic spectra in a range of
expected effective temperatures ($4000$ to $6000 \mathrm{K}$), surface gravities ($0.8$ to $4$~dex),
and metallicities (from $0.0$ to $-1.5~\mathrm{dex}$). The synthetic spectra were created
with the Kurucz atmospheric models and {\em synthe} code
\citep{castelli,synthe,klinux} and with the {\tt GALA} \citep{gala} visual inspector
{\em sline}. Clean, unblended lines were selected and their laboratory
wavelengths were taken from VALD3 \citep{vald3}. We ran {\tt DAOSPEC} through
the {\tt DOOp} self-configuration wrapper \citep{2014A&A...562A..10C} leaving the RV
totally free, and then created a cleaned linelist by selecting only lines that
were measured in at least three spectra. A second run of {\tt DOOp}/{\tt
DAOSPEC} with the cleaned linelist produced the final observed RVs. 

In a few cases {\tt DAOSPEC} crashed, and there were spectra with discrepant 
parameters\footnote{Roughly $10$ to $15$\% of the spectra
had RV spread or FWHM quite different from the typical values, or had RV values
outside the range $-200 \le RV \le +450$\,\kps.}. All these spectra were visually inspected and
--- when appropriate --- re-run imposing a starting RV within $10$~\kps
around the values obtained for other spectra of the same star.
The vast majority of the problematic cases happened with LR2 spectra, as expected.
%%%%
{While we treated the UVES RED--L and RED--U spectra separately 
($\overline{\Delta \mathrm{RV}_\mathrm{UL}} = 0.22 \pm 0.77$~\kps, for 201 spectra),
in the following sections and for method to method comparison purposes, both determinations 
corrected for the shift of the telluric lines were averaged to 
provide one single value per epoch.}
%%%%

\subsection{Mask cross-correlation with {\tt iSpec}\label{sec:ispec}}

{\tt iSpec} is an integrated spectroscopic software framework with the
necessary functions for the measurement of radial velocities, the determination 
of atmospheric parameters and individual chemical abundances \citep{2014A&A...569A.111B}. 
{\tt iSpec} includes several observed and synthetic
masks and templates for different spectral types that
can be used to derive radial velocities by cross-correlation. 
However, very few of them cover the full wavelength range 
of our target spectra. For this 
work we chose to cross-correlate the LR2, HR21 and UVES spectra (UVES RED--L and RED--U 
spectra treated independently), with a mask built on the atlas of the Sun 
published by \cite{2000vnia.book.....H} and which ranges from $373$ nm to $930$~nm.
The cross-correlation of a target spectrum with that mask provides a velocity profile, 
the peak of which is fitted with a second order polynomial. {\tt iSpec} also provides 
the RV error and FWHM of the correlation function fitted by a Gaussian. Due to the large 
number of spectra we are dealing with in this study, it is not possible to check each 
cross-correlation visually. We thus developed an automatic processing pipeline which 
can identify the most uncertain determinations of RV.

For each target spectrum, two coarse estimates of RV were first obtained over 
a large interval, from $-$500 to $+$500~\kps, with a step of 5~\kps\,and with a step of 1~\kps. 
If the two resulting values differ significantly, this is an indication of 
a poor cross-correlation. After several tests, we found that it was the most efficient 
way to detect unreliable determinations which can be due to fast rotation, bad SNR, 
binarity, or spectral mismatch. Such bad determinations were rejected. They 
represent $\sim$1\% of the LR2 and HR21 determinations, but 8\% and 17\% of 
the UVES spectra covering the shorter and longer wavelength part, respectively. 
For well behaving cross-correlations a more precise RV was determined by using a step 
of 0.5~\kps\,over an interval of $\pm$50~\kps\,around the first velocity estimation.
{When both determinations are available, the UVES RED--L and RED--U RV 
values ($\overline{\Delta \mathrm{RV}_\mathrm{UL}} = 0.16 \pm 0.46$~\kps, for 187 spectra)
were averaged to provide one single measurement per epoch. 
}
%%%

\section{Results\label{sec:results}}

\subsection{Comparison between the different methods\label{sec:comparison}}

\begin{figure*}[htbp]
\includegraphics[]{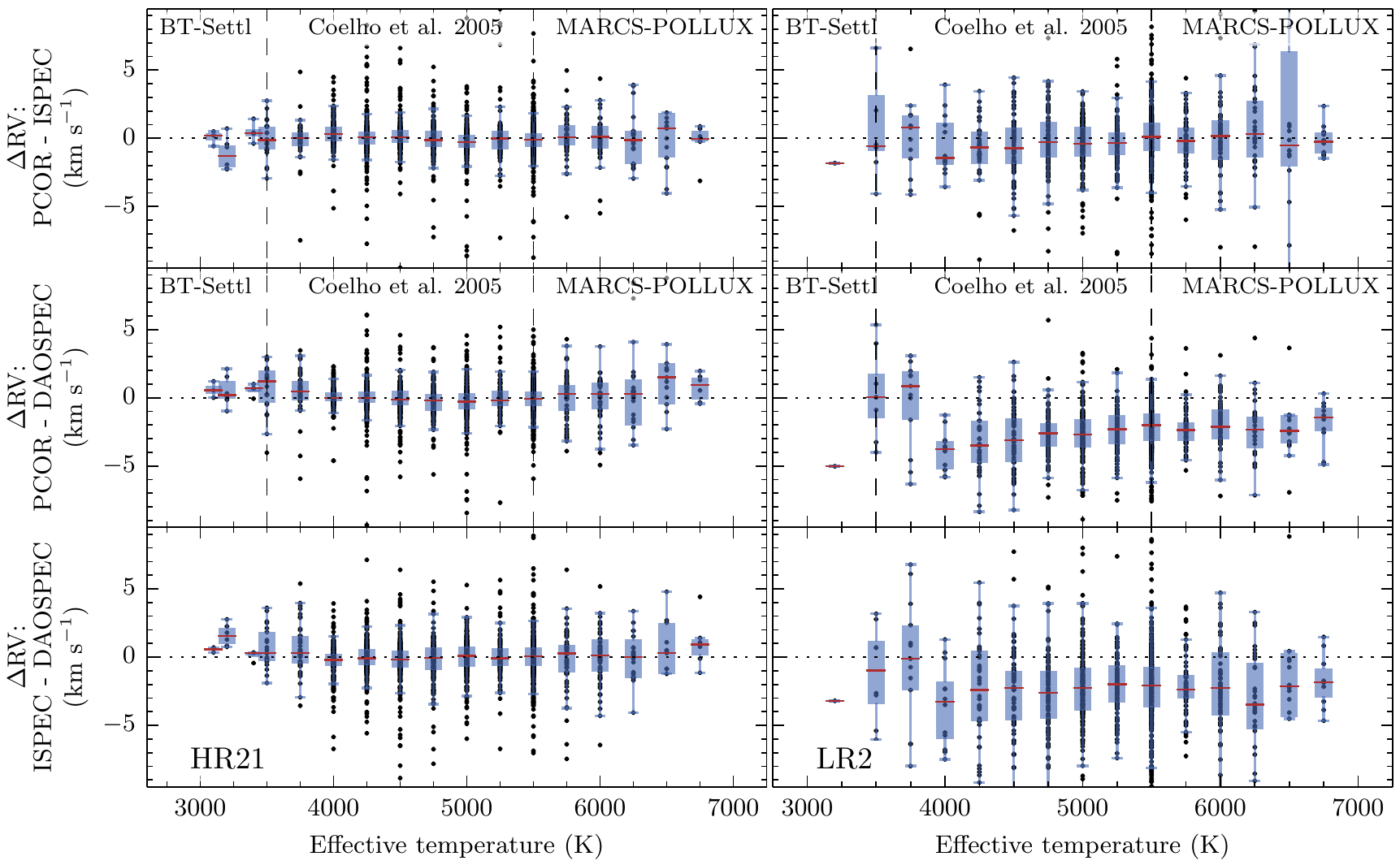}
\caption[]{Method to method comparisons for the HR21 (left panel) and LR2 (right panel) data:
RV differences are plotted as a function of the effective temperature (dark blue points).
The blue boxes extend from the lower to
the upper quartile, while the whiskers cover the range of values without outliers
from which the median was computed (red horizontal bar).
\label{fig:comparison:giraffeTHRLR}}
\end{figure*}

\begin{figure}[htbp]
\includegraphics[]{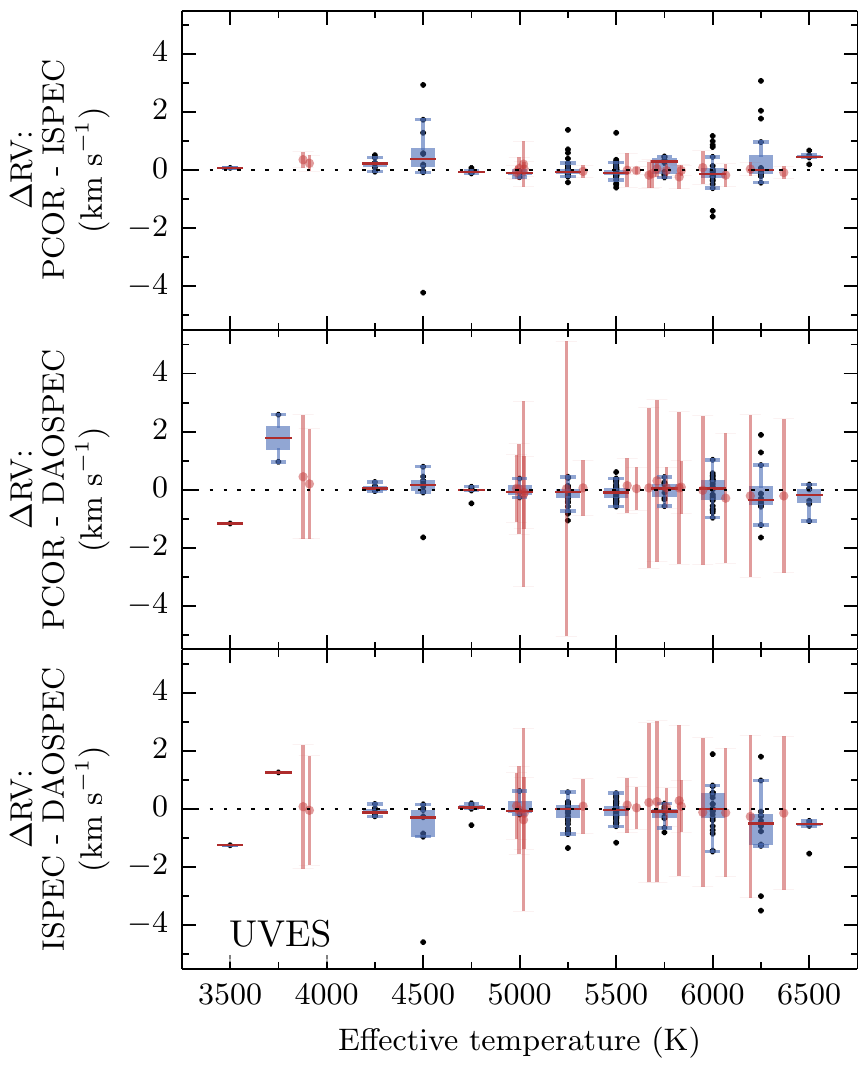}
\caption[]{Method to method comparisons for the UVES data: 
RV differences are plotted against the effective temperature.
The blue points are the
measurements obtained for the SEP targets, while the red disks represent the Gaia RV standards. 
The blue boxes extend from the lower to
the upper quartile, while the whiskers cover the range of values without outliers
from which the median was computed (red horizontal bar). The \teff\ is taken from the
results of the template optimisation with {\tt PCOR} (Sect.~\ref{sec:templateoptimization}).
\label{fig:comparison:uvesT}}
\end{figure}

The radial velocities derived by the three methods were compared
with one another, and the variation of their relative deviations,
$\Delta\mathrm{RV}$, is plotted as a function of \teff\, (obtained in
Sect.~\ref{sec:templateoptimization}) in Figs.~\ref{fig:comparison:giraffeTHRLR} and \ref{fig:comparison:uvesT} 
for the HR21 and LR2, and UVES data, respectively. The comparison was made for all spectra with
SNR$\ge$5. Because in the HR21 and LR2 wavelength ranges hot
stars have too few available spectral lines, often broadened due
to higher rotation, and blended to strong hydrogen lines, we decided
not to use the {\tt iSpec} (Sect.~\ref{sec:ispec}) and {\tt DAOSPEC} (Sect.~\ref{sec:daospec})
methods above 7000~K and therefore limited our comparison to
the cooler stars which represent 99\% of our sample.

\begin{table}[htbp]
\caption{\label{tab:biases} Method to method biases derived per \teff\, range}
\center
\begin{tabular}{rrrr}
\hline\hline
\teff~range & $\overline{\Delta \mathrm{RV}}$ & $\sigma$ & $\overline{\Delta \mathrm{RV}}$/$\sigma$ \\
(K) & (\kps) & (\kps) & \\
\hline
\multicolumn{4}{c}{LR2}\\
\multicolumn{4}{c}{{\tt PCOR} vs {\tt iSpec}}\\
3500 - 5500 & $+$0.020 & 0.134 & 0.150 \\
5500 - 8000 & $-$0.060 & 0.239 & 0.251 \\
\multicolumn{4}{c}{{\tt PCOR} vs {\tt DAOSPEC}}\\
3500 - 5500 & $-$2.108 & 0.103 & 20.452 \\
5500 - 8000 & $-$2.316 & 0.169 & 13.691 \\
\multicolumn{4}{c}{{\tt iSpec} vs {\tt DAOSPEC}}\\
3500 - 8000 & $-$2.208 & 0.154 & 14.374 \\
\hline
\multicolumn{4}{c}{HR21}\\
\multicolumn{4}{c}{{\tt PCOR} vs {\tt iSpec}}\\
$<$ 3500 & $-$0.410 & 0.370 & 1.108 \\ 
3500 - 5500 & $-$0.040 & 0.032 & 1.268 \\
5500 - 8000 & $+$0.100 & 0.148 & 0.674 \\
\multicolumn{4}{c}{{\tt PCOR} vs {\tt DAOSPEC}}\\
$<$ 3500 & $+$0.431 & 0.288 & 1.494 \\ 
3500 - 5500 & $-$0.077 & 0.034 & 2.289 \\
5500 - 8000 & $+$0.298 & 0.167 & 1.780 \\
\multicolumn{4}{c}{{\tt iSpec} vs {\tt DAOSPEC}}\\
3500 - 8000 & $-$0.032 & 0.039 & 0.812 \\
\hline
\multicolumn{4}{c}{UVES}\\
\multicolumn{4}{c}{{\tt PCOR} vs {\tt iSpec}}\\
3250 - 8000 & $-$0.050 & 0.034 & 1.492 \\
\multicolumn{4}{c}{{\tt PCOR} vs {\tt DAOSPEC}}\\
3250 - 8000 & $-$0.056 & 0.034 & 1.650 \\
\multicolumn{4}{c}{{\tt iSpec} vs {\tt DAOSPEC}}\\
3250 - 8000 & $-$0.040 & 0.054 & 0.737 \\
\hline
\end{tabular}
\end{table}

The method to method offset or bias, $\Delta\mathrm{RV}$, was estimated
from the median of the differences. As various libraries of synthetic
spectra (BT-Setl, Coelho et al. 2005, MARCS) were used
with the PCOR procedure to cover the full \teff\, range (Sect.~\ref{sec:correlation by templates}),
the offset of the method relatively to the other ones was derived,
where possible, for each grid separately.

The determinations are listed in Table~\ref{tab:biases} with their \teff\, range,
median bias ($\overline{\Delta \mathrm{RV}}$), corresponding error ($\sigma$), and the ratio between
the bias and the error. The latter value being used to assess
the bias significance using a two-tailed test \citep[eq. 20,][]{2014A&A...562A..97D}. 
At the 1\% significance level, it implies for what follows
that an offset between two collections of measurements will be
judged significant if the corresponding $\overline{\Delta \mathrm{RV}}$/$\sigma$ > 2.57.

In Fig.\ref{fig:comparison:giraffeTHRLR}, the method to method differences show a larger scatter in
the LR2 wavelength domain than in the HR21 with a slight trend
toward 4000~K, its absolute value tending to increase with decreasing
temperature. This should be related to the nature of the
methods and to their behaviour toward line blending (stronger
in cooler stars) and resolving power (lower in LR2). The
dispersion of the RV
differences between methods varies from 0.34~\kps\,for UVES
data to 2.24~\kps\,in the LR2 results, while the bias is lower
than 0.5~\kps\,except in LR2 where it peaks at 2.32~\kps\,in
absolute value. At the 1\% significance level, if the bias-to-error
ratio given in Table~\ref{tab:biases} is larger than 2.57, the measurements are
definitely biased. We will therefore consider that all three methods
are providing consistent results, except in LR2 where, according
to its limits of applicability (see Sect.~\ref{sec:daospec}), measurements
obtained by {\tt DAOSPEC} show a systematic bias relative to {\tt iSpec}
and {\tt PCOR}. LR2 is indeed a region with a weaker signal and a lot
of heavily blended spectral atomic and molecular lines that may
bias their individual localisation. As a consequence,
the LR2 results obtained by {\tt DAOSPEC} were not considered in what follows.

\onllongtab{
\onecolumn
\tabcolsep=0.1cm
\begin{longtab}
\begin{longtable}{rlrrrrrrrrrr}
\caption{\label{tab:rvepoch} Epoch radial velocity determinations. Observations with SNR $<$ 5 are shown
in italics for information purposes only, but are not further taken into account in the discussion.
{The table provides: the stars' EID, the considered observing mode, the SNR, the heliocentric Julian date, 
the final de-trended multi-epoch barycentric RV determinations for each method as well as their 
weighted mean and corresponding error bar ($\sigma_{\rm RV}$). (The complete version of the table will be made available at the CDS.)}
}\\
\hline\hline
 &  &  &  & \multicolumn{2}{c}{{\tt ISPEC}} & \multicolumn{2}{c}{{\tt XCOR.D}} & \multicolumn{2}{c}{{\tt DAOSPEC}} & \multicolumn{2}{c}{{\tt W.Mean}} \\
EID & Reg. & SNR & HJD & RV & $\sigma_{\rm RV}$ & RV & $\sigma_{\rm RV}$ & RV & $\sigma_{\rm RV}$ & RV & $\sigma_{\rm RV}$ \\
    &      &     &     & (\kps) &  (\kps) & (\kps) &  (\kps) & (\kps) &  (\kps) & (\kps) &  (\kps) \\
\hline
\endfirsthead
\caption{continued.}\\
\hline\hline
 &  &  &  & \multicolumn{2}{c}{{\tt ISPEC}} & \multicolumn{2}{c}{{\tt XCOR.D}} & \multicolumn{2}{c}{{\tt DAOSPEC}} & \multicolumn{2}{c}{{\tt W.Mean}} \\
EID & Reg. & SNR & HJD & RV & $\sigma_{\rm RV}$ & RV & $\sigma_{\rm RV}$ & RV & $\sigma_{\rm RV}$ & RV & $\sigma_{\rm RV}$ \\
    &      &     &     & (\kps) &  (\kps) & (\kps) &  (\kps) & (\kps) &  (\kps) & (\kps) &  (\kps) \\
\hline
\endhead
\hline
\endfoot
{\it        21404 }&{\it  HR21 }&{\it            1 }&{\it 55271.64688 }&{\it     297.13 }&{\it       0.23 }&{\it      16.24 }&{\it       3.52 }&{\it     -15.92 }&{\it       3.69 }&{\it       0.16 }&{\it       2.55}\\
{\it        21404 }&{\it  HR21 }&{\it            4 }&{\it 55285.57863 }&{\it     -24.10 }&{\it       1.32 }&{\it       6.59 }&{\it       1.26 }&{\it       7.43 }&{\it       1.15 }&{\it       7.01 }&{\it       0.85}\\
       21404 &  HR21 &           11 & 55519.85760 &       8.04 &       0.74 &       6.73 &       0.50 &       8.00 &       1.14 &       8.02 &       0.68\\
       21404 &  HR21 &           19 & 55521.75806 &       6.96 &       0.77 &       5.28 &       0.36 &       5.37 &       0.54 &       5.32 &       0.32\\
{\it        21404 }&{\it   LR2 }&{\it            3 }&{\it 55521.83296 }&{\it      -4.38 }&{\it       1.00 }&{\it       9.39 }&{\it       1.63 }&{\it -- }&{\it --  }&{\it       2.50 }&{\it       0.96}\\
{\it        21404 }&{\it   LR2 }&{\it            2 }&{\it 55522.78352 }&{\it       2.89 }&{\it       0.67 }&{\it      10.26 }&{\it       1.78 }&{\it -- }&{\it --  }&{\it       6.58 }&{\it       0.95}\\
       21404 &   LR2 &            9 & 55592.59354 &       4.46 &       2.60 &       4.55 &       0.40 & -- & --  &       4.50 &       1.31\\
{\it        21498 }&{\it  HR21 }&{\it            1 }&{\it 55271.64688 }&{\it     299.45 }&{\it       0.23 }&{\it     561.31 }&{\it       8.77 }&{\it     216.03 }&{\it      12.49 }&{\it     257.74 }&{\it       6.25}\\
{\it        21498 }&{\it  HR21 }&{\it            2 }&{\it 55285.57863 }&{\it     194.84 }&{\it       0.31 }&{\it     494.15 }&{\it       7.15 }&{\it     -20.88 }&{\it       1.65 }&{\it     222.70 }&{\it       2.45}\\
       21498 &  HR21 &           10 & 55519.85760 &     300.37 &       0.40 &     296.64 &       0.81 &     249.11 &       1.32 &     298.51 &       0.45\\
\end{longtable}
\end{longtab}

}

\begin{figure}[htbp]
\includegraphics[]{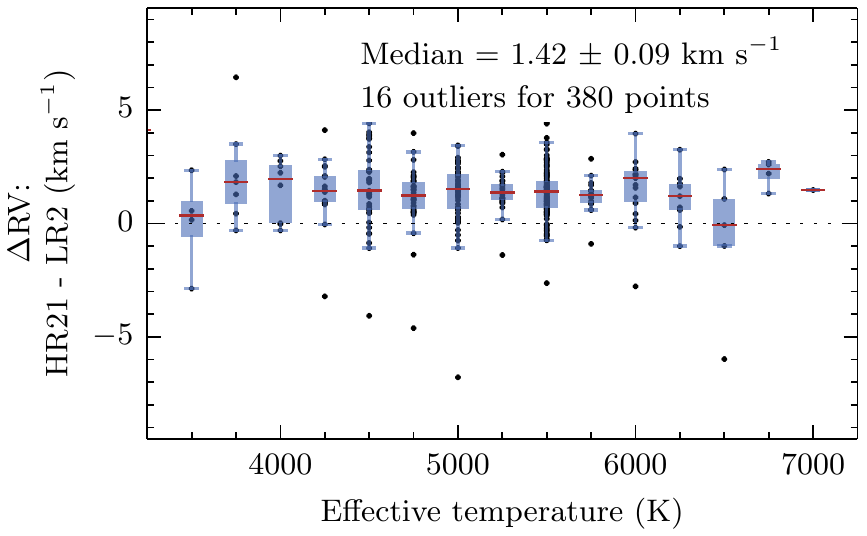}
\caption[]{Determination of the RV bias between HR21 and LR2: 
Differences between the star per star RV median derived in HR21 and LR2
are plotted against the effective temperature (dark blue points).
The blue boxes extend from the lower to
the upper quartile, while the whiskers cover the range of values without outliers
from which the median was computed (red horizontal bar). The median of the
differences, its standard deviation, and the number of outliers according to 
Chauvenet's criterion are provided.
\label{fig:comparison:HR21vsLR2}}
\end{figure}

\begin{figure}[htbp]
\includegraphics[]{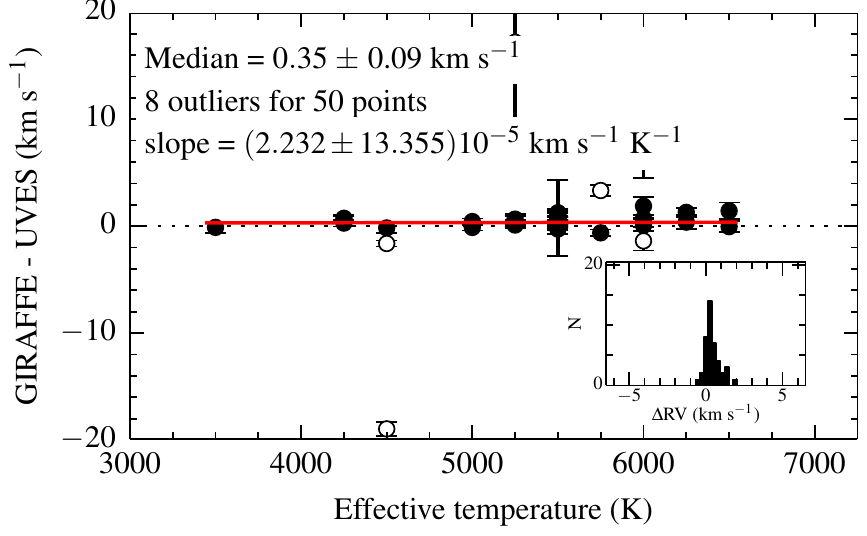}
\caption[]{Determination of the RV offset between combined GIRAFFE and UVES measurements:
RV differences are plotted as a function of the effective temperature. Open disks
are outliers. The median of the differences and the corresponding standard deviation are provided, as
well as the slope of the best fit drawn (red line) through the points.
Height outliers were filtered out by applying Chauvenet's criterion.\label{fig:comparison:GvU}}
\end{figure}

The combined RV determinations of 380 stars observed in
both the HR21 and LR2 domains and with combined errors less
than 2~\kps\,were used to estimate the bias between the 2 GIRAFFE
setups as plotted in Fig.~\ref{fig:comparison:HR21vsLR2}. To construct this 
graph all
significant method measurements available for one star in one
of the 2 domains were combined into a median. The datasets
are significantly biased by 1.42~\kps, with a dispersion 
of 1.12~\kps. Because the number of stars that each
were observed in LR2 and UVES is too small (i.e., 14) and that
these spectra have a lower SNR, we first placed the LR2 measurements
in the HR21 RV-scale. Then, for those stars having
a combined RV error smaller than 1~\kps, the comparison between
GIRAFFE and UVES data is shown in Fig.~\ref{fig:comparison:GvU} and provides
a median offset of 0.35~\kps\,for a semi-interquantile dispersion
of 0.40~\kps. This offset is of the same order and sign
as the median telluric-line shift ($\rm{0.49} \pm \rm{0.74}$~\kps) we found
in Sect.~\ref{sect:reduction_spec}, and of the same order as those found in the Gaia
ESO survey \citep[e.g, lower panel of Fig.~5 in][]{2014A&A...565A.113S} with other GIRAFFE observing
modes and using a different reduction pipeline. It can therefore
be considered as not directly related to the data calibration issues.
In both cases (HR21 vs LR2 and GIRAFFE vs UVES), no
significant global temperature dependence was found. We subtracted
the bias from each individual GIRAFFE measurement.
Then, after having filtered out the outliers by applying the Chauvenet criterion in
3 successive iterations, we have combined
the two or three RV measurements by computing their weighted
mean and corresponding uncertainties.

The results are stored in Table~\ref{tab:rvepoch}, with the stars' EID,
the considered observing mode, the SNR, the heliocentric Julian date, 
the final de-trended multi-epoch barycentric RV determinations for each
method, as well as their weighted mean and corresponding 
error bar ($\sigma_{\rm RV}$).

\subsection{Comparison with the catalogue of radial velocity standard stars for Gaia\label{sec:comparisonwcatalogue}}

\begin{figure}[htbp]
\includegraphics[]{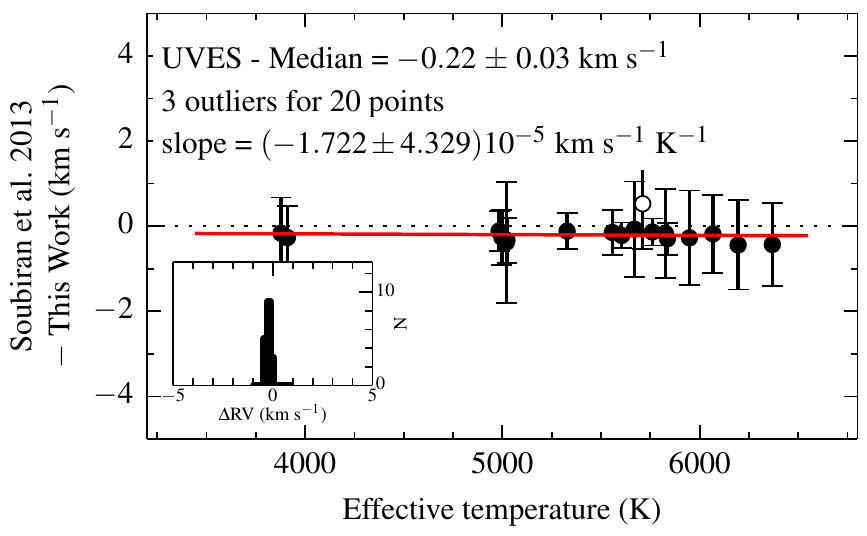}
\caption[]{Gaia RV standard stars: comparison between UVES RV measurements (this work) 
and those found in the \citet{2013A&A...552A..64S} catalogue.
The median of the differences and the corresponding standard deviation are
provided, as well as the slope of the best line drawn (red line) through
the points. The histogram distribution of the deviations are provided in
the Figure inset. Outliers were filtered out by applying Chauvenet's criterion.
\label{fig:comparison:STD}}
\end{figure}

We have corrected all the measurements for the internal biases and brought our
radial velocities on the same scale, but we still need to check how the results
compare to the framework defined by the RV standard stars compiled by the
IAU Commission 30.
Therefore, in the ESO archive, we retrieved 
76 UVES spectra of 20 targets in common with the catalogue of RV standard stars for 
Gaia \citep{2013A&A...552A..64S}. We adopted the 
astrophysical parameters proposed in the PASTEL database \citep{2010A&A...515A.111S}
and applied to these data the same procedures and methods as above 
(see RV results in Fig.~\ref{fig:comparison:uvesT}, red disks).
The comparison of these results with the measurements found in the catalogue 
is shown in Fig.~\ref{fig:comparison:STD}.
UVES radial velocities appear to be systematically overestimated 
by $\rm{0.22}$~\kps\,for an inter-quantile dispersion of 0.12~\kps, which is a known 
offset for UVES as it is of the same order and sign as the median 
deviation ($8 \pm 17$~m\AA\, redwards of 5200~\AA) obtained by \citet{2003A&A...407.1157H} 
between the sky line positions in the UVES and Keck atlases. Although it should be accounted for 
when compared to other catalogues, we prefer to let the user decide to apply the correction
and to not include it in our tabulated values. 

\subsection{Multi-epoch analysis\label{sec:multiepoch}}

\begin{figure}[htbp]
\includegraphics[]{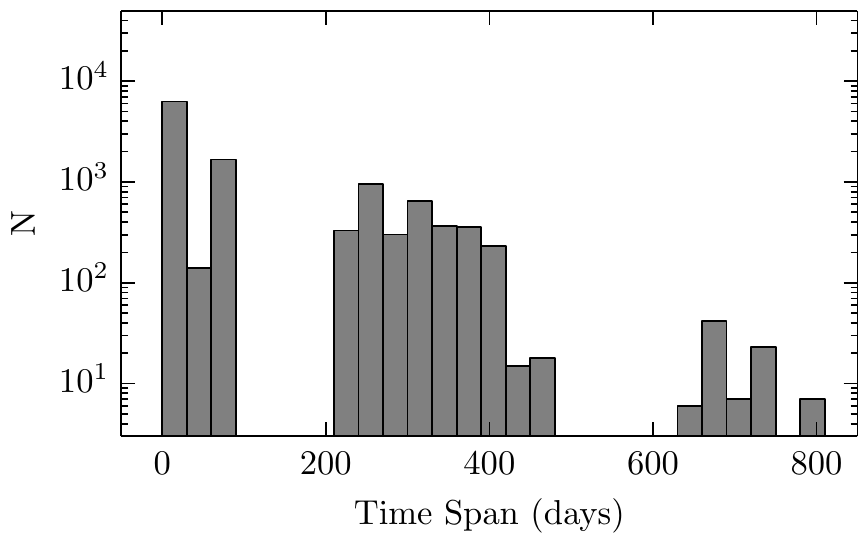}
\caption[]{Distribution of the time span between 2 observations of a same target.\label{fig:tspan}}
\end{figure}

Most targets were observed at least twice a night, then later during
the other periods. Fig.~\ref{fig:tspan} shows the distribution of the time
span between 2 spectra of the same star. Our widest time coverage
(808 days) was obtained for target EID\,80185.

For a given star, we combined all individual epoch measurements
from all methods into one single value. In order to take into
account the RV scatter --- which may not be due only to random
errors --- in the final error bar estimate, we assumed that
the errors of each measurement follows a normal distribution.
Then we applied a Monte-Carlo scheme with 1000 realisations
per determination, and we computed for each star the median
and corresponding semi-interquantile dispersion.

\begin{figure}[htbp]
\includegraphics[]{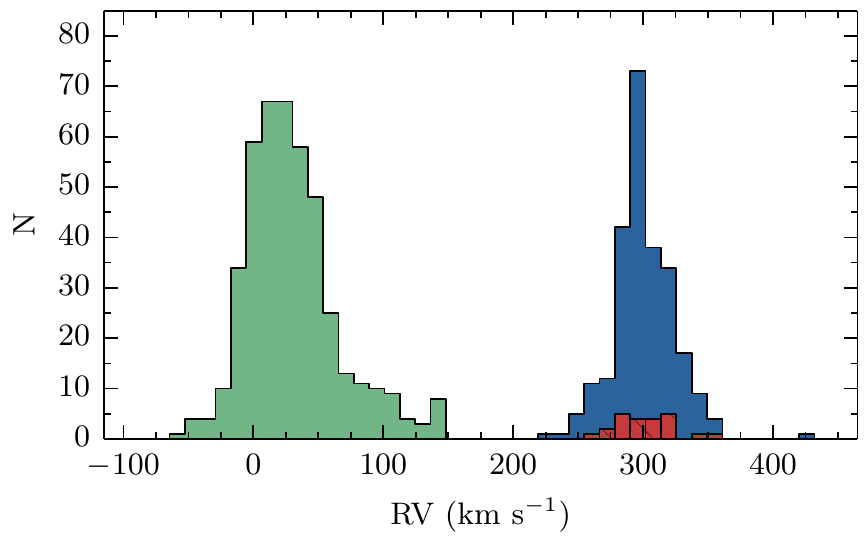}
\caption[]{Median RV distribution: Stars with RVs lower or equal than 200~\kps~are counted in the
green bins, other targets fall in the blue ones. C and S stars are identified by red 
hatched bins.\label{fig:rvdist}}
\end{figure}

The distribution of the combined RVs is shown in Fig.~\ref{fig:rvdist}. It
features two main stellar populations belonging to the LMC and
to the Milky Way (MW), with radial velocities larger or smaller
than 200~\kps, respectively. The results are stored in Table~\ref{tab:rvintegrated},
which provides the star’s EID, the median RV ($\overline{RV}$), its error bar
($\sigma_{\mathrm{\overline{RV}}}$), the total number of individual measurements used to compute
the median (n), the number of epochs (N), and the time span
of the observations. For the stars found to have variable photometry
by OGLE, Table~\ref{tab:rvintegrated} also gives the corresponding period of
photometric variation and the OGLE variability flags \citep{2012AcA....62..219S} in the last columns.

A significant part of the stars in our sample show small to large amplitude 
radial velocity scatter. To identify the most constant ones,
we estimated, method per method, the consistency of the 
epoch RV measurements ($RV_i$ and corresponding error $\sigma_i$) 
and of their weighted mean, $\left\langle{RV}\right\rangle$, we computed 
the scatter as follows

\begin{equation}
\chi^2 = \sum_i \left(\left\langle{RV}\right\rangle - RV_i\right)^2 / \sigma_i^2
\end{equation}

\noindent as well as the associated probability, $P(\chi^2)$, that the deviations
are within the known error boundaries. From the method by
method analysis of these deviations, we deduce that 34 out of
the 489 stars with more than 2 measurements have $\mathrm{P(}\chi^\mathrm{2}\mathrm{)} < 10^\mathrm{-6}$
and show significant RV variations whatever the method used.
Those stars are noted ''VAR'' in Col.~(7) of Table~\ref{tab:rvintegrated}. Additionally,
44 targets ranked ''VAR'' by at least 2 methods or with 
$\mathrm{P(}\chi^\mathrm{2}\mathrm{)} < 0.00135$ for all 3 methods are labelled ''VAR?''. 
Conversely, stars
having $P(\chi^2) \ge 0.00135$, more than 2 measurements spread over
a time span larger than 180 days, with $\sigma_{\rm \overline{RV}} <$~1~\kps, and
which therefore does not show any significant RV scatter in the
results of all 3 methods are noted ''RV-REF'' in Table~\ref{tab:rvintegrated} (represents
145 stars). If one of the criteria above, on time span or on
$\sigma_{\rm \overline{RV}}$, is not fulfilled, then the target received the ''RV-REF?'' label
(represents 125 stars). For the remaining 141 objects (of the
489), nothing conclusive can be inferred.

\onllongtab{
\onecolumn
\tabcolsep=0.1cm
\begin{longtab}
\begin{longtable}{rrrrrrrrr}
\caption{\label{tab:rvintegrated} Median radial velocities, LMC membership, and variability status:
{
The table provides the star’s EID, the median RV ($\overline{RV}$), its error bar
($\sigma_{\mathrm{\overline{RV}}}$), the total number of individual measurements used to compute
the median (n), the number of epochs (N), and the time span of the observations. When useful, 
we provide in col.~(7) a few remarks in the form of labels: LMC (bona fide LMC member),
LMC? (possible LMC member), VAR (RV variable), VAR? (prossible RV variable), 
RV-REF (target with constant RV over the time span), RV-REF? (possibly constant in RV),
\ion{Ca}{ii} K em. (shows emission in the \ion{Ca}{ii} K line), 
\ion{Ca}{ii} T em. (shows emission in the \ion{Ca}{ii} triplet), and
\ion{H}{i} em. (shows emission in the Paschen lines).
For the stars found to have variable photometry by OGLE, we further give in the last columns the corresponding 
period of photometric variation and the OGLE variability flags \citep{2012AcA....62..219S}. (The complete version of the table will be made available at the CDS.)
}
}\\
\hline\hline
 & & & & & & & \multicolumn{2}{c}{OGLE}\\
\cline{8-9}
 EID & $\overline{RV}$ & $\sigma_{\mathrm{\overline{RV}}}$ & $n$ & $N$ & Time Span & Remark(s) & Period & VAR flag \\
  &  (\kps) & (\kps) &  &  & (days) &  & (days) & \\
\hline
\endfirsthead
\caption{continued.}\\
\hline\hline
 & & & & & & & \multicolumn{2}{c}{OGLE}\\
\cline{8-9}
 EID & $\overline{RV}$ & $\sigma_{\mathrm{\overline{RV}}}$ & $n$ & $N$ & Time Span & Remark(s) & Period & VAR flag \\
  &  (\kps) & (\kps) &  &  & (days) &  & (days) & \\
\hline
\endhead
\hline
\endfoot
%%
% Id: UPDATE_RV_INTEG.py 1983 2016-07-24 17:27:33Z yves.fremat 
%%
% SNR >= 5
   21404 &     6.16 &     0.46 &   8 &   3 &    72.74 & RV-REF? &   &   \\
   21498 &   297.53 &     0.78 &  12 &   5 &    72.74 & VAR &   &   \\
   22969 & {\it   308.19 }&{\it      0.17} &   5 &   5 &   320.95 & LMC, C Star & 51.17 & LPV/OSARG \\
   23916 &    72.48 &     1.25 &   2 &   1 &     0.00 &  &   &   \\
\end{longtable}
\end{longtab}
\label{page:end}

}

\section{Discussion\label{sec:discussion}}

\subsection{Astrophysical parameters}

\begin{figure}[htbp]
\includegraphics[]{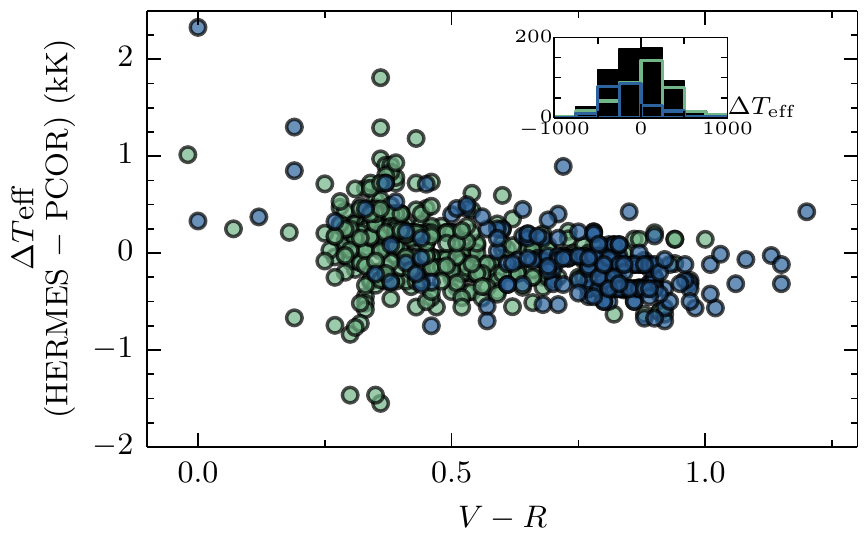}
\caption[]{Effective temperature deviations versus $V-R$: Differences are between 
the effective temperature of the best HERMES template
and the one derived during synthetic template optimisation.
Stars with RV < 200~\kps\,are shown in green, while the other ones are
in blue. The histogram of the deviations is given in the Figure inset, with the
same colour coding and with the complete distribution shown in black. 
\label{fig:teffsyntvshermes}}
\end{figure}

The astrophysical parameters of the best matching synthetic
spectrum are by-products of the template optimisation scheme
adopted with the {\tt PCOR} program (Sect.~\ref{sec:templateoptimization}). 
To have an estimate
of their consistency level, we compared the effective temperatures
obtained with this procedure to those found in the literature
of the best corresponding matching HERMES spectrum
(Sect.~\ref{sec:APhermes}). The differences between the former and
latter determinations are plotted in Fig.~\ref{fig:teffsyntvshermes} in function 
of the $V-R$ colour. Distinction is made between stars with RV smaller and larger 
than 200~\kps\,(see Sect.~\ref{sec:lmc}). If we except a few
outliers, all points fall within 1000~K, with a median value and
dispersion of $30 \pm 335 \mathrm{K}$ (black histogram in
inset of Fig.~\ref{fig:teffsyntvshermes}). Although the distribution looks symmetric, deviations
follow a trend with $V-R$: stars with $\mathrm{RV} \ge 200$~\kps\,
showing on average an offset of $-120 \pm 260 \mathrm{K}$, the other ones
have differences of the order of $140 \pm 340 \mathrm{K}$. Such behaviour
may have various origins. It can be due to the non-completeness
of the libraries (e.g., in terms of \feh), to the model
assumptions (e.g., 1D atmosphere models for red giants) or to
the accuracy of the atomic data used to compute the spectra.
Anyhow, we generally have a fair agreement between
the observed and final synthetic spectra, and --- as
will be shown hereafter --- we have a good consistency between
the resulting APs (\teff, \logg, and \feh) and the different stellar
populations to be expected in the MW and LMC. A lower
estimate of the template mismatch effects and of the parameter
errors on the RV determinations can be deduced from Fig.~1 in
\citet{2014A&A...562A..97D}. In our temperature range and assuming deviations
of 500~K, this would lead to biases of the order of 200~\mps, which is of the 
same order as our smallest RV error bars.

\subsection{Variability\label{sec:variability}}

It is reasonable to suspect that part of the detected RV variability
(Sect.~\ref{sec:multiepoch}) is due to stellar multiplicity, to pulsations or to any
other astrophysical phenomenon at the origin of RV jitter. Unfortunately,
we currently do not have a sufficient number of observations to model the variations
and to derive non degenerate solutions. A few targets,
however, show clear spectroscopic evidence for a companion.
Seven have confirmed RV variability, while we lack sufficient data for the other two
(EID 50823 and EID 98406). In Table~\ref{tab:rvintegrated}, we labelled ''SB2'' 
or ''SB2?'' these targets with composite spectra.

\begin{figure}[ht]
\includegraphics[]{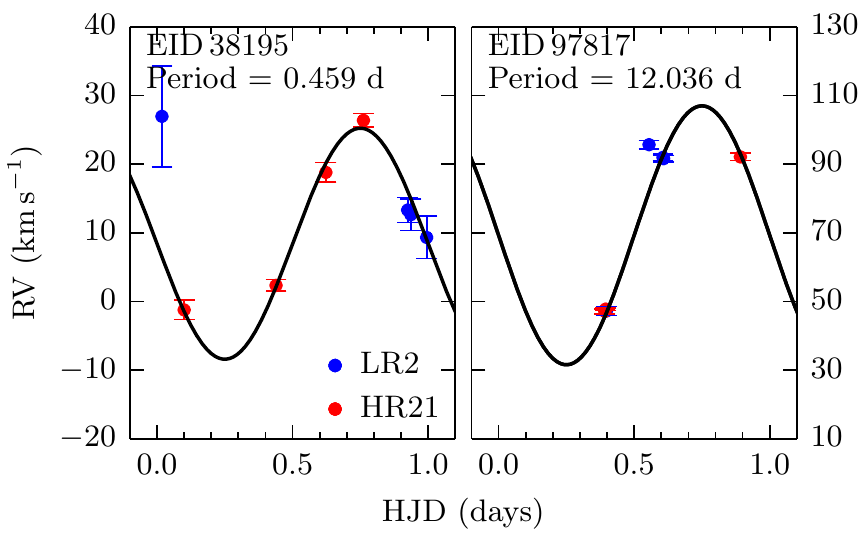}
\caption[]{Phase diagrams of the eclipsing EID\,38195 (left panel) and
SB1 EID\,97817 (right panel) binaries\label{fig:rvd}}
\end{figure}

\begin{figure}[htbp]
\includegraphics[]{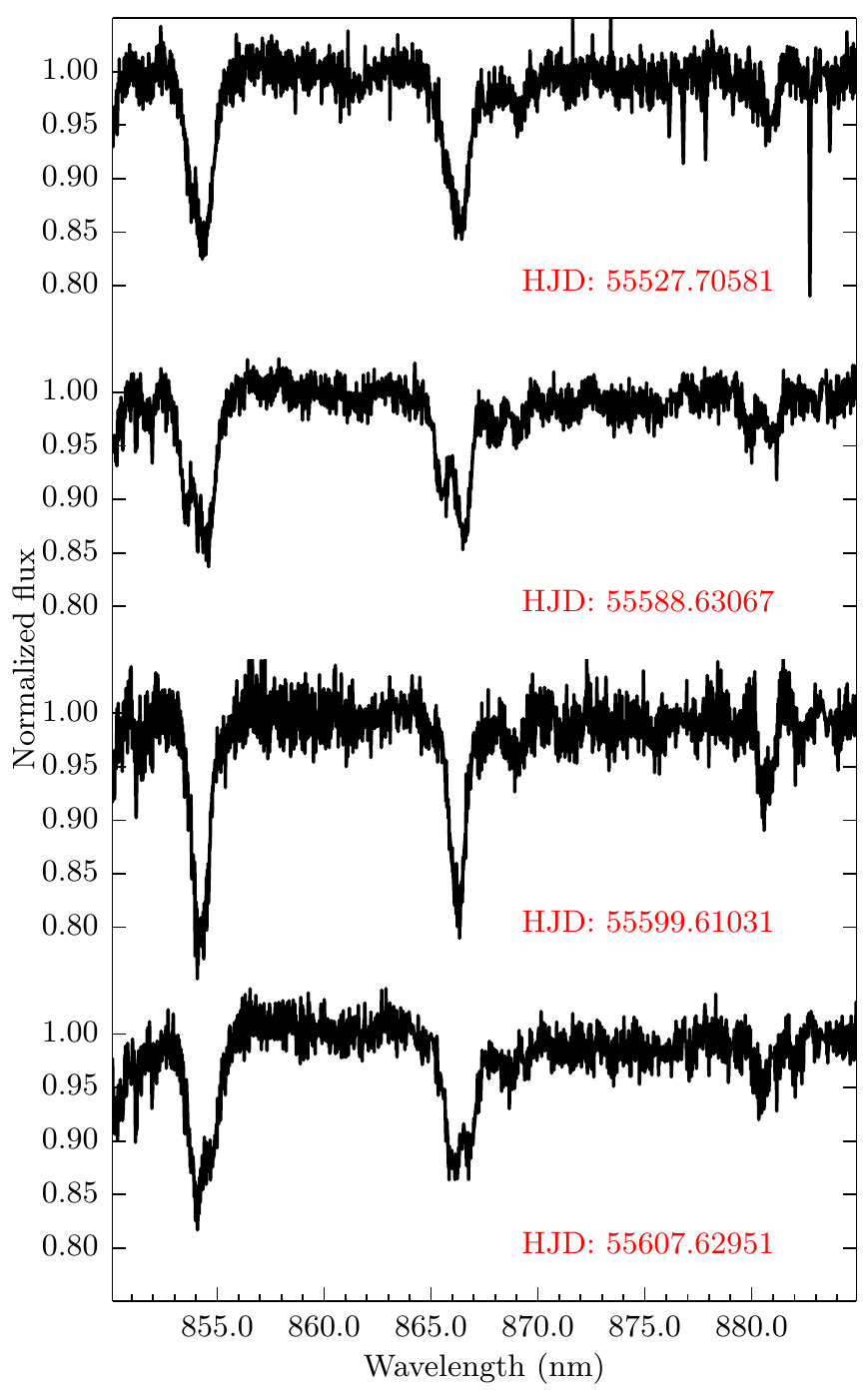}
\caption[]{Spectroscopic variations of the eclipsing SB2 EID\,84084.\label{fig:84084}}
\end{figure}

Among the 5 eclipsing binaries identified by OGLE, 3 are
also SB2. It is the case of EID\,84084, for which we plot in
Fig.~\ref{fig:84084} the HR21 spectroscopic variability. EID\,38195 is detected
VAR, but its variations cannot be directly phased
with the photometric period of 0.45~d. We therefore submitted
its RVs to a multi-step RV-curve fitting procedure developed by
\citet{2012ocpd.conf...71D} and found a close solution at P = 0.459~d
(Fig.~\ref{fig:rvd}, left panel).We only have two LR2 epoch measurements
for EID\,72521, which therefore was not classified VAR by our
procedure even though the difference between the 2 RVs is significant
(i.e., $P(\chi^2) < 10^\mathrm{-6}$).

\begin{figure}[ht]
\includegraphics[]{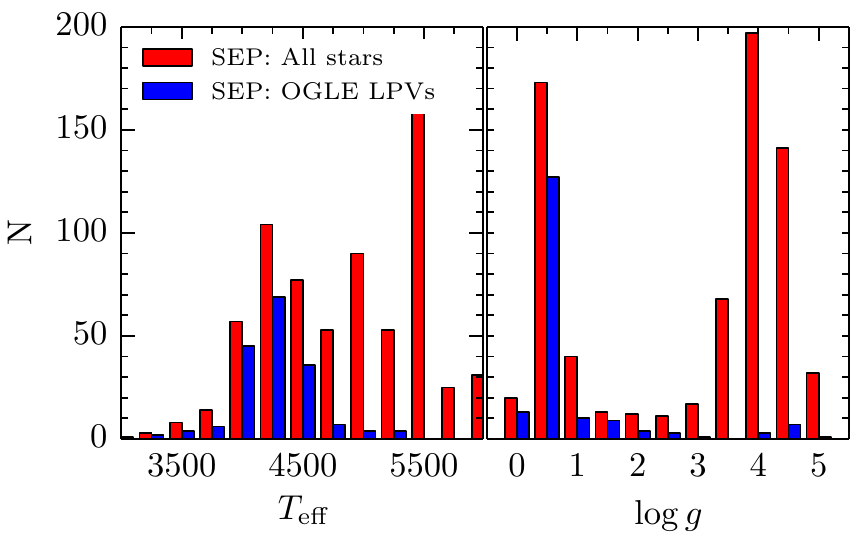}
\caption[]{Distribution of \teff\, and \logg\, among SEP and LPV stars\label{fig:lpv}}
\end{figure}

Four stars have ellipsoidal variations in the OGLE data. Two
(EID\,30581 is VAR and EID\,96413 is VAR?) are spectroscopically
variable. EID\,107847 does not show any significant RV variation
over a period of time of 385 days in the 7 UVES spectra, while
the OGLE period is 302 days. EID\,92169 was observed 9 times
over 347.92 days. It exhibits a larger RV scatter ($\sim 3$~\kps),
but this is not sufficient for it to be detected as variable by applying
our various criteria.

Among the identified VAR and VAR? stars, 17 show photometric
long period variability in the OGLE catalogue as would
be expected from evolved cool stars. If one excepts EID\,97817,
all these stars have a \logg~$< 2.5$ and have RVs compatible with
the LMC. More generally, as shown in Fig.~\ref{fig:lpv}, the majority of
stars flagged ''LPV'' by OGLE have \teff\, and \logg\, consistent
with those found for red giants. Because EID\,97817 exhibits
quite a high amplitude RV scatter and astrophysical parameters
that do not belong to an evolved cool star, we suspect it to be
a SB1 rather than an LPV. As a matter of fact, the orbit fitting procedure
of \citet{2012ocpd.conf...71D} proposes a solution at the exact
same OGLE period (Fig.~\ref{fig:rvd}, right panel).

EID\,50856 is noted by OGLE to be a galactic RR Lyrae,
which agrees with its magnitude ($V = 16.37$) and with the astrophysical
parameters of the nearest synthetic spectrum (\teff~$=6250 \mathrm{K}$,
\logg~$=4.0$, and \feh~$=-1.5$). Two of the RV methods
found it variable in RV. The rather high median RV (i.e., $420.43 \pm 1.53$~\kps) 
and its low \feh\, suggests that it is a halo population II star. 
%It is worth noting that several
%LMC RR Lyrae were already detected by Gaia \citep{2016CoKon.105....3C} in 
%the SEP and had their light curves published on ESA’s 
%website\footnote{\url{http://www.cosmos.esa.int/web/gaia/iow_20150305}}.

\subsection{LMC membership\label{sec:lmc}}

\begin{figure}[htbp]
\includegraphics[]{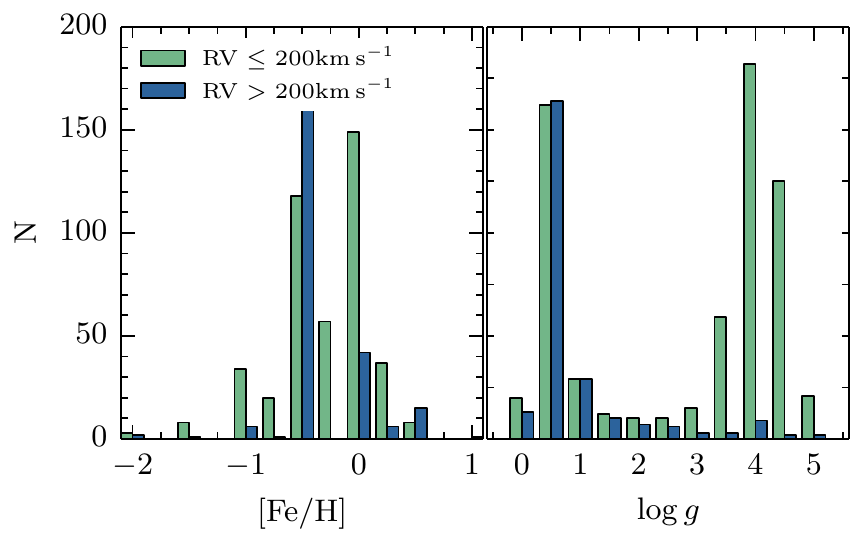}
\caption[]{\feh\, and \logg\, distributions of the SEP targets
\label{fig:metdist}}
\end{figure}

The one-square-degree SEP field covers a small part of the LMC.
Therefore a significant number of stars in our sample are LMC
members, with radial velocities different from those found on
average in the MW. Indeed we find a double-peaked RV distribution
(Fig.~\ref{fig:rvdist}) with one top at $\sim 24$~\kps\,and another
at $\sim 298$~\kps. As shown by the \logg\, and \feh\, estimates
(Fig.~\ref{fig:metdist}), stars with
RV~$\ge$~200~\kps\,further tend to be \feh\, depleted \citep[e.g.,][]{1992A&A...266...85T} and have
a surface gravity lower than 2.5 as we expect for the brightest
members of the LMC. In our sample, 203 targets with radial
velocities larger than 200~\kps, \logg~$< 2.5$, and metallicities
lower than in the Sun can therefore be regarded as bona fide
LMC members. In addition, 51 more stars have RVs consistent with
the LMC but higher metallicities or higher \logg. Among the 178
targets with OGLE LPV variability flagged, 169 belong to these
LMC or candidate LMC stars, which do confirm their red giant
status. Both categories are respectively labelled ''LMC'' and ''LMC?''
in Table~\ref{tab:rvintegrated} and are highlighted in Fig.~\ref{fig:pointings}. LMC stars appear to be
randomly distributed over the SEP field, and follow no obvious
structure.

\subsection{Targets of particular interest\label{sec:peculiar}}

\begin{figure*}[htbp]
\includegraphics[]{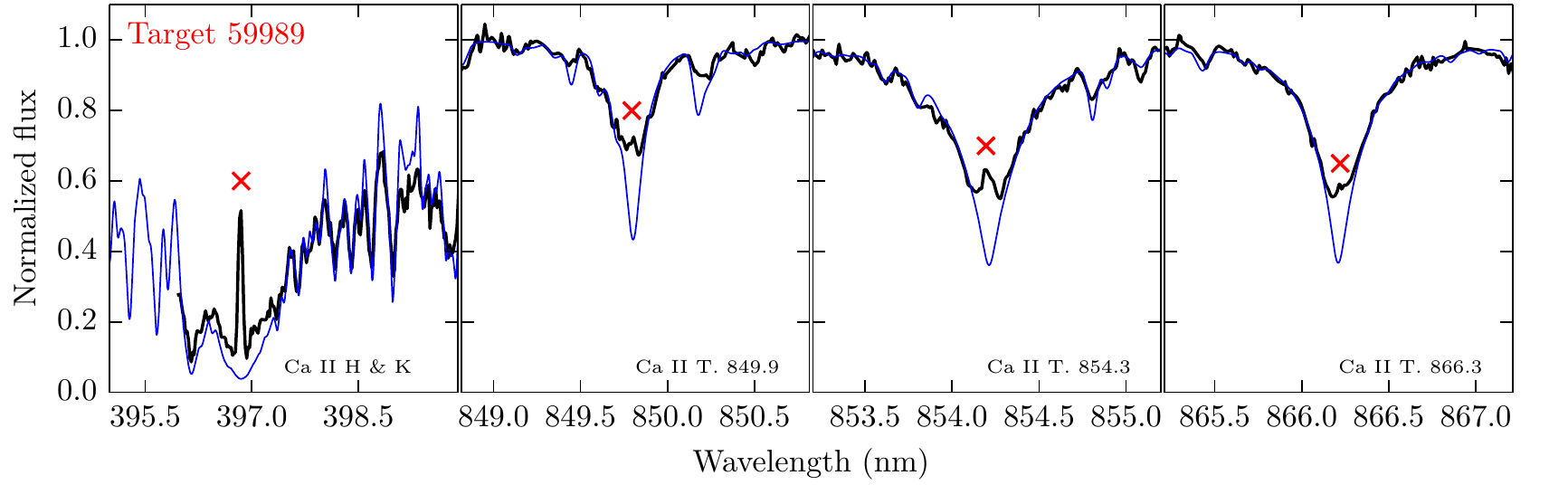}
\caption[]{Comparison between the best fitted template (dashed blue curve) and the observed LR2 and HR21 data of 
EID\,59989 (solid curve). Signs of chromospherical activity are detected in the core of the calcium lines (red 
crosses).\label{fig:59989}}
\end{figure*}

Nine G and K-type SEP stars show emission-like features, sometimes
variable, in the near-IR Calcium triplet and/or in the blue
\ion{Ca}{ii} K line (see Fig.~\ref{fig:59989}). These stars, which generally have
\logg$\ge 4$ (exceptions are EID\,77381 and EID\,102310) are tagged
''\ion{Ca}{ii} K and T em.'' or ''\ion{Ca}{ii} K em.'' in Table~\ref{tab:rvintegrated} and probably
are chromospherically active. Among these, EID\,88013 shows
periodic photometric variability due to spots. In addition, 1 B-type
supergiant, EID\,68904, exhibits emission in its hydrogen
Paschen lines (''\ion{H}{i} em.'' in Table~\ref{tab:rvintegrated}).

The comparison of the GIRAFFE spectra with HERMES
data (Sect.~\ref{sec:APhermes}), led us to find a few stars with features only seen
in evolved objects and which we identified as carbon (18 targets) and
S stars (6 targets). All these objects have RVs typical of those
found in the LMC. The C stars further already have an entry in
the carbon star catalogue of \citet{2001A&A...369..932K}. As we do
not have any mask nor synthetic spectrum adapted to C stars, we
derived their radial velocity using high SNR data of VY
UMa.

\begin{figure*}[htbp]
\includegraphics[]{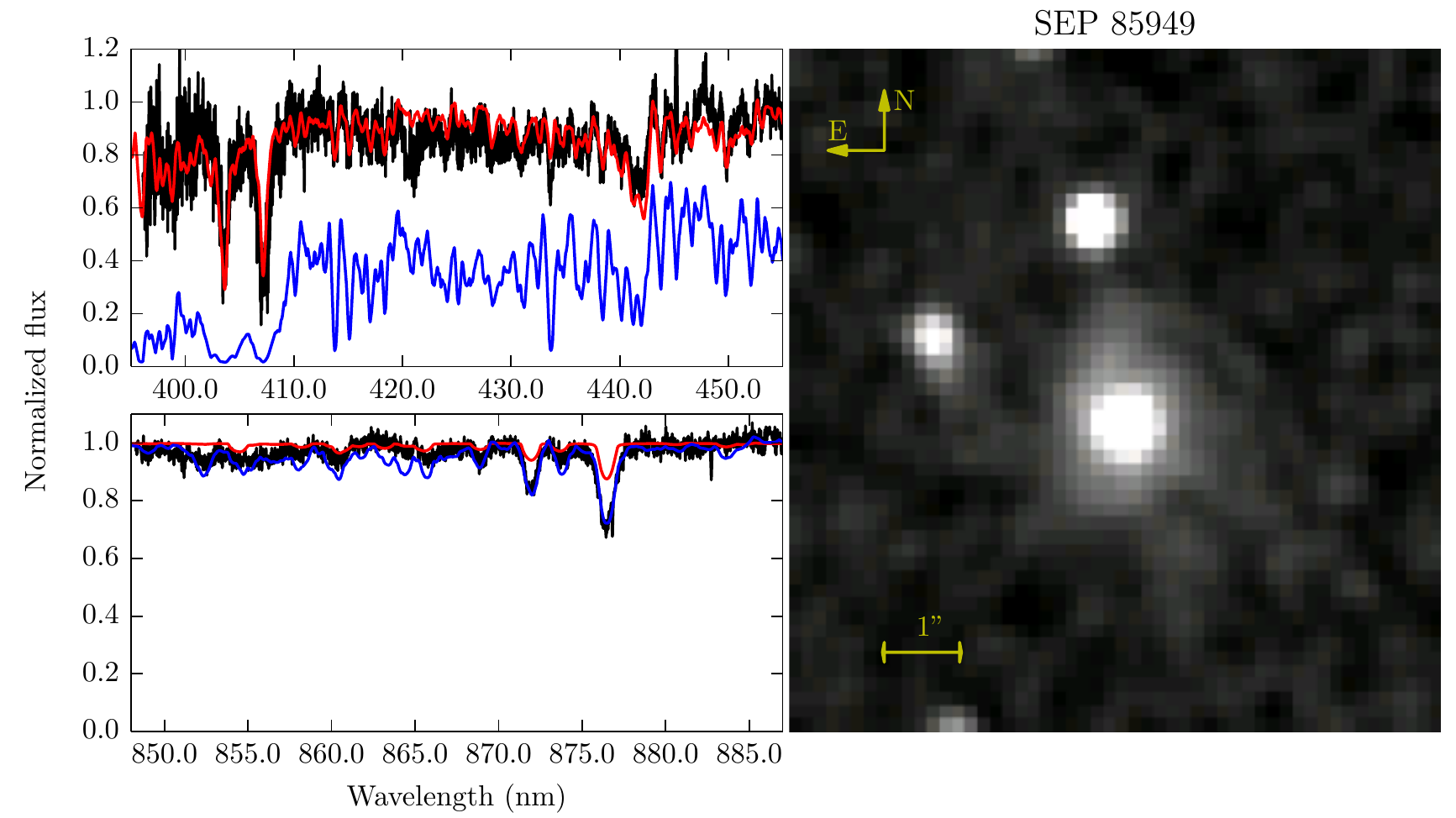}
\caption[]{The observed spectra (black curve) of the Galaxy EID\,85949 is compared to
2 theoretical stellar spectra computed with the same \teff\, and \logg\, (4000~K and 1.00, respectively), 
but having a different metallicity. The blue line spectrum has a Solar-like metallicity, while 
the red one is depleted by $-$2.5~dex. Both spectra are shifted 
by $\mathrm{RV} = 7810$~\kps\,($z = 0.026$) to match the 
observations.\label{fig:85949}}
\end{figure*}

Due to the random fibre allocation, we could identify one galaxy (Fig.~\ref{fig:85949}) 
already mentioned in {\it the 2M++ galaxy redshift catalogue} \citep{2011MNRAS.416.2840L}. 
By fitting the corresponding averaged LR2 and HR21 data with synthetic spectra, 
we estimated a redshift of 0.026.

%radial velocity of 7810~\kps~which is close to the published value, 7837~\kps. 

\section{Conclusions\label{sec:conclusions}}

Among the 747 targets selected at random in the 1-square-degree
field centred around the South Ecliptic Pole and observed
with FLAMES, 725 had spectra with $\mathrm{SNR} \ge 5$. By visually inspecting
these and by performing a systematic comparison with
HERMES observations, we identified one galaxy, 18 C stars, and
6 new S stars, while the use of various libraries of synthetic spectra
provided us with a first stellar classification.
As a main result, we measured the radial velocities on all the
FLAMES (UVES and GIRAFFE) data by applying 3 different
methods in order to assess their zero-point and precision. From
the RV, \feh, and \logg\, distributions we have extracted 203
objects that are bona fide LMC members, and 51 that have RVs
and \feh\, or \logg\, values compatible with those we expect
for the LMC in the considered magnitude ranges. Multi-epoch
observations enabled us to identify 78 RV variable stars as well
as to highlight those targets with the most stable radial velocity
(145 stars). Seven confirmed SB2s --- among which 3 are
eclipsing --- and 2 candidate SB2 stars have composite spectra.

In Gaia commissioning, the satellite had several periods
of Ecliptic Pole Scanning Law (EPSL), during which the satellite repeatedly
scanned the North and South Ecliptic Poles (i.e.
each pole was observed twice\footnote{twice for the 2 Gaia telescopes}) every 6 hours. These periods
allowed Gaia’s Radial Velocity Spectrometer \citep[RVS,][]{2004MNRAS.354.1223K,2011EAS....45..181C}
to collect a large number of observations
of the GIRAFFE and UVES stars presented in this
article. These were used to validate the RVS ground-based processing
pipeline \citep{2011EAS....45..189K} and to make a first appraisal
of its radial velocity performance for the faint-magnitude stars
\citep{cropper2014,seabroke2015}. Now that the routine
mission is on-going, the SEP GIRAFFE-UVES sample is part of
the nominal RVS ground-based processing pipeline. It is used to
monitor and assess the convergence of the RVS performance in
the faint star regime.

\begin{acknowledgement}
We thank Dr J.R. Lewis for his careful reading of the manuscript and for his
constructive comments.
This research is supported in part by ESA-Belspo PRODEX funds, in particular 
via contract n$^{\rm o}$~4000110150/4000110152 “Gaia early mission Belgian consolidation”,
and by the German Space Agency (DLR)
on behalf of the German Ministry of Economy and Technology via Grant 50 QG 1401.
PJ acknowledges financial support by the European Union FP7 programme 
through ERC grant number 320360.
{UH and TN acknowledge support from the Swedish National Space Board (Rymdstyrelsen).}
This research was achieved using the POLLUX database (\url{http://pollux.graal.univ-montp2.fr})
operated at LUPM  (Université Montpellier II - CNRS, France with the support of the PNPS and INSU,
and has made use of the SIMBAD database \citep{2000A&AS..143....9W}, operated at CDS, Strasbourg, France.
We would like to thank R.Napiwotzki for sharing with us his {\sc uvbybeta} computer code.
Figure~\ref{fig:pointings} was generated using the Kapteyn python package available from \url{https://www.astro.rug.nl/software/kapteyn/}.
\end{acknowledgement}

\bibliographystyle{aa} % style aa.bst
\bibliography{AA_2016_29549_R1} % your references Yourfile.bib

\vfill

\fbox{\parbox{0.9\textwidth}{Tables found in pages \pageref{page:start} to \pageref{page:end} are 
available in the electronic edition of the journal at \url{http://www.aanda.org} as well as at the 
CDS in electronic format.} }

\onecolumn

\Online

\end{document}